\documentclass[aps,prl,superscriptaddress,twocolumn,longbibliography]{revtex4-1}
\usepackage{bbm}
\usepackage{mathrsfs}
\usepackage{amsmath}
\usepackage{amsfonts}
\usepackage[colorlinks=true,linkcolor=blue,urlcolor=blue,citecolor=blue,anchorcolor=blue]{hyperref}
\usepackage{graphicx,epstopdf}
\usepackage{subfigure}
\usepackage{epsfig}
\usepackage{dcolumn}
\usepackage{bm}
\usepackage{color}
\usepackage{natbib}
\usepackage{amssymb}
\usepackage{xcolor}
\usepackage{braket}
\usepackage{ulem}
\usepackage{float}
\usepackage{braket}
\usepackage{changes}
\allowdisplaybreaks[4]

\begin{document}
	
\title{Superradiance enhances and suppresses fermionic pairing based on universal critical scaling in two order parameters systems}
\author{Yilun Xu}
\email{xuyilun@alumni.pku.edu.cn}
\affiliation{State Key Laboratory for Mesoscopic Physics, School of Physics, Frontiers Science Center for Nano-optoelectronics, Peking University, Beijing 100871, China}
\affiliation{Beijing Academy of Quantum Information Sciences, Beijing 100193, China}

\begin{abstract}
	Distinguished from the system with one order parameter, systems described by two or more order parameters will manifest more complex and much richer phase diagram and critical phenomena. In systems of two order parameters, the phase transition of one order parameter may influence the strength of another. Focus on the Landau's theory of continuous phase transitions, we give a general physcial quantity to decide the changing rate of the two order parameters based on a general formula of free energy. Taking two-mode Rabi model and the 1D Fermi Dicke model as the examples, we verify our analytical results and show how the superradiant phase transition manipulates the two-spin pairing strength and the superconductor band gap. Our work proposes the new paradigm to study the complex systems with two or more order parameters and provides novel avenue to enhancing or suppressing the desired physical effect by such interplay. 
\end{abstract} 

\maketitle
\textit{Introduction.} In order to understand the classic condensed matter systems such as classic Ising ferromagnetic model~\cite{Peierls1936} and classic XY model~\cite{PhysRevLett.20.589,AIZENMAN1980281}, Ginzburg–Landau theory (also called Landau's theory)~\cite{HOHENBERG20151} has always served as a powerful tool to depict the physical properties and the phase transtions in these systems, especially the scaling behaviors around the critical temperature $T=T_c$. Although it will be invalid to deal with strong correlated systems, whose quantum fluctuation is nonnegligible, it still provides a concise avenue to study the quantum systems in classic oscillator limit~\cite{PhysRevLett.115.180404,PhysRevLett.117.123602,PhysRevLett.119.220601} or thermodynamic limit~\cite{PhysRevLett.104.130401,PhysRevLett.122.193201,PhysRevLett.120.183603,PhysRevLett.112.173601,PhysRevLett.108.043003,PhysRevLett.124.073602,PhysRevLett.132.073602} and instruct the experimental development of quantum simulation~\cite{RN105,RN103,RN104,PhysRevLett.105.043001,PhysRevLett.104.130401,PhysRevLett.92.130403,RN107,annurev,PhysRevLett.112.143002,PhysRevLett.112.143003,PhysRevLett.112.143004,doi:10.1126/science.abd4385,PhysRevLett.133.173602}.



The basic idea of Landau's theory is to express the free energy as the function of "macroscopic" order parameters such as photon number, magnetism, disorders. "macroscopic" means the value of the order parameters is typically much larger than the quantum fluctuation, with the quantum effect effectively suppressed. By minimizing the free energy of the system, one can obtain the order parameters of the ground states or the stable states, revealing the basic property of the system. Remarkably, the emerging (from zero to nonzero) of certain order parameter refers to spontaneous symmetry breaking, such as $Z_2$ symmetry in Dicke model~\cite{PhysRevA.7.831,PhysRevE.67.066203,HEPP1973360} and the global $O(2)$ symmetry in classic XY model, and $U(1)$ symmetry in isotropic Heisenberg XY model~\cite{PhysRevLett.20.589}, indicating the phase transition occurs.

In general, the system described by one order parameter will give us only the normal and ordered phase, corresponding to the zero and nonzero order parameter. But when it comes to systems with two or more order parameters, not only the phase diagrams become much richer and more complex, but also the interplays of these order parameters are interesting and potentially useful. The order parameters may reinforce or suppress each other in different phases, which provides us more possibility to implement the quantum simulation and manipulation. In particular, we are dedicated to enhancing certain quantum effect, such as the Cooper pair condense, which is originated from the superconducting band gap according to Bardeen-Cooper-Schrieffer theory~\cite{PhysRev.108.1175}. Instead of searching new systems with larger superconductor band gap, we can modify the band gap of the present system through manipulating transitions of other physcial observables.

In this letter, concentrate on continuous phase transition, we first give the universal critical scaling rate to evaluate the interplay of the two order parameters right beyond the critical boundary. It facilitates us to determine the property of the system in ordered phase, only making use of the knowledge on critical points. Next, we take two examples to show how it provides precise judgement on how the superradiant phase transition affects the fermionic pairing. On one hand, we design a two mode Rabi model through the floquet engineering, where the two spin pairing strength can be effectively modified by the superradiance in one mode. On the other hand, we propose the one-dimensional Fermi Dicke model with attractive interactive, where the superradiance from the continous phase transition will always suppress the superconducting band gap. Although the 1D system is only viewed as the toy model, it still serves as a good example to unveal the interplays of the two order parameters.

\textit{Universal critical scaling rate of two order parameters.} Consider a system described by two macroscopic order parameters $\left(O_1,O_2\right)$ with the environmental parameter vector $\vec{\lambda}$, and the free energy gives as
\begin{align}\label{free_energy}
	F=F(O_1,O_2;\vec{\lambda})=\sum_{v_1,v_2}g_{v_1,v_2}(\vec{\lambda})O_1^{v_1}O_2^{v_2}.
\end{align}
Generally, the order parameter $O_i$ is well selected as nonzero real numbers, such as the number of photon in the superradiant phase transition or the mean magnetism in the ferromagnetic phase. And we assume that all the coefficients are the continuous and analytical function of the environmental parameter $\vec{\lambda}$, and the second order coefficients $g_{2,0}(\vec{\lambda}),g_{0,2}(\vec{\lambda})\neq0$ with the change of $\vec{\lambda}$. In other words, the situation $g_{2,0}(\vec{\lambda})g_{0,2}(\vec{\lambda})=0,~\forall\vec{\lambda}$ is excluded. The system status will be determined by the order parameters $\left(\bar{O}_1(\vec{\lambda}),\bar{O}_2(\vec{\lambda})\right)$, where the free energy is minimum, i.e., it satisfies
\begin{align}
	F(\bar{O}_1(\vec{\lambda}),\bar{O}_2(\vec{\lambda}))=\min_{(O_1,O_2)}F(O_1,O_2;\vec{\lambda}).
\end{align}
Notice that the system status $\left(\bar{O}_1(\vec{\lambda}),\bar{O}_2(\vec{\lambda})\right)$ are the functions of the environmental parameter $\vec{\lambda}$. Without loss of generality, we assume it keeps the $Z_2$ symmetry as $F(O_1,O_2;\vec{\lambda})=F(O_1,-O_2;\vec{\lambda})$. Then we define that the 2nd order critical point $\vec{\lambda}_c$ seperates the normal phase (NP) with $\bar{O}_2=0$ and the ordered phase with $\bar{O}_2\neq0$. Consider the the system enters into the ordered phase from the critical point by changing the environmental parameter $\vec{\lambda}_c\to\vec{\lambda}_c+\delta\vec{\lambda}$, defining $\bar{O}_i^c\equiv\bar{O}_i(\vec{\lambda}_c)$.

\begin{widetext}
	In this case, the critical boundary is given by $\partial_2^2F(\bar{O}_1^c,0;\vec{\lambda}_c)=0$, 
	\begin{align}\label{delta_F0}
		\delta F\equiv F(O_1,O_2;\vec{\lambda})-F(\bar{O}_1^c,0;\vec{\lambda})=\sum_{i}\partial_iF(\bar{O}_1^c,0;\vec{\lambda})\delta O_i+\dfrac{1}{2}\sum_{i,j}\partial_i\partial_jF(\bar{O}_1^c,0;\vec{\lambda})\delta O_i\delta O_j+\mathcal{O}[(\delta O_i)^3].
	\end{align}
	Here, $\delta O_i(\vec{\lambda})\equiv \bar{O}_i(\vec{\lambda})-\bar{O}_i^c$. In the following, we omit the position of differentiation $(\bar{O}_1^c,0)$ for simply writing. Consider the $Z_2$ symmetry of $O_2$, the equation $\partial_1^n\partial_2F=0,~\forall n$ is always satisfied. We obtain $\partial_iF(\vec{\lambda}_c)=0$ to recover the ground state $\delta O_i=0$. Set $\vec{\lambda}=\vec{\lambda}_c+\delta\vec{\lambda}$, we can calculate the minimum value $\left(\delta O_1,\delta O_2\right)$ by asking $\dfrac{\partial\delta F(\vec{\lambda}_c+\delta\vec{\lambda})}{\delta\delta O_i}=0$ as follows
	
	\begin{align}\label{solve_min1}
		\begin{cases}
			0=\dfrac{\partial\delta F}{\partial\delta O_1}=\partial_1F(\vec{\lambda}_c+\delta\vec{\lambda})+\partial_1^2F(\vec{\lambda}_c+\delta\vec{\lambda})\delta O_1+\dfrac{\partial_1\partial_2^{v_{m1}}F(\vec{\lambda}_c+\delta\vec{\lambda})(\delta O_2)^{v_{m1}}}{v_{m1}!}+\mathcal{O}[\dots]\\
			0=\dfrac{\partial\delta F}{\partial\delta O_2}=\partial_2^2F(\vec{\lambda}_c+\delta\vec{\lambda})\delta O_2+\partial_1\partial_2^{v_{m1}}F(\vec{\lambda}_c+\delta\vec{\lambda})\delta O_1(\delta O_2)^{v_{m1}-1}+\dfrac{\partial_2^{v_{m2}}F(\vec{\lambda}_c+\delta\vec{\lambda})(\delta O_2)^{v_{m2}-1}}{(v_{m2}-1)!}+\mathcal{O}[\dots]
		\end{cases}.
	\end{align}
\end{widetext}

Here we define $v_{m1}$ and $v_{m_2}$ as $v_{m1}\equiv\min\left\{v~|~\sum_{v_1}v_1\bar{O}_1^{v_1-1}g_{v_1,v}(\vec{\lambda}_c)\neq0\right\}$, and $v_{m2}\equiv\min\left\{v~|~\sum_{v_1}\bar{O}_1^{v_1}g_{v_1,v}(\vec{\lambda}_c)\neq0\right\}$. We can expand $\partial_2^2F$ at $\lambda_c$ as $\partial_2^2F(\vec{\lambda}_c+\delta\vec{\lambda})\sim\mathcal{O}(\left|\delta\vec{\lambda}\right|)<0$, driving the $\bar{O}_2$ becomes finite according to the second equation. Meanwhile, we assume the special case that $\partial_1F(\vec{\lambda}_c+\delta\vec{\lambda})=0$, which means that $\delta\vec{\lambda}$ is the environmental pertubation bringing no influence to the stability of $\bar{O}_1^c$ directly. Based on this assumption, the first equation tells the scaling rate between $\delta O_1$ and $\delta O_2$, which reads as 
\begin{align}\label{interplay}
	\delta O_1\approx-\dfrac{\partial_1\partial_2^{v_{m1}}F(\bar{O}_1^c,0;\vec{\lambda}_c)}{v_{m1}!\partial_1^2F(\bar{O}_1^c,0;\vec{\lambda}_c)}(\delta O_2)^{v_{m1}},
\end{align}
with $\delta\vec{\lambda}\to0$. Notice the denominator is positive according to the stability, we can see that the sign of the term $\partial_1\partial_2^{v_{m1}}F(\bar{O}_1,0;\vec{\lambda}_c)$ decides whether the $\bar{O}_1$ increases or decreases right after the phase transition. Although Eq.~(\ref{interplay}) seems difficult to give us an explicit physical insight to a concrete system, it tends to provide a good approach to evaluate the property of the phase only based on the differentiation on the critical line, saving the cost of calculation. Furthermore, we take two concrete examples to clarify the application of Eq.~(\ref{interplay}) in the following.

\begin{figure}[tb]
	\centering
	\hspace{-0.8cm}
	\includegraphics[width=0.52\textwidth]{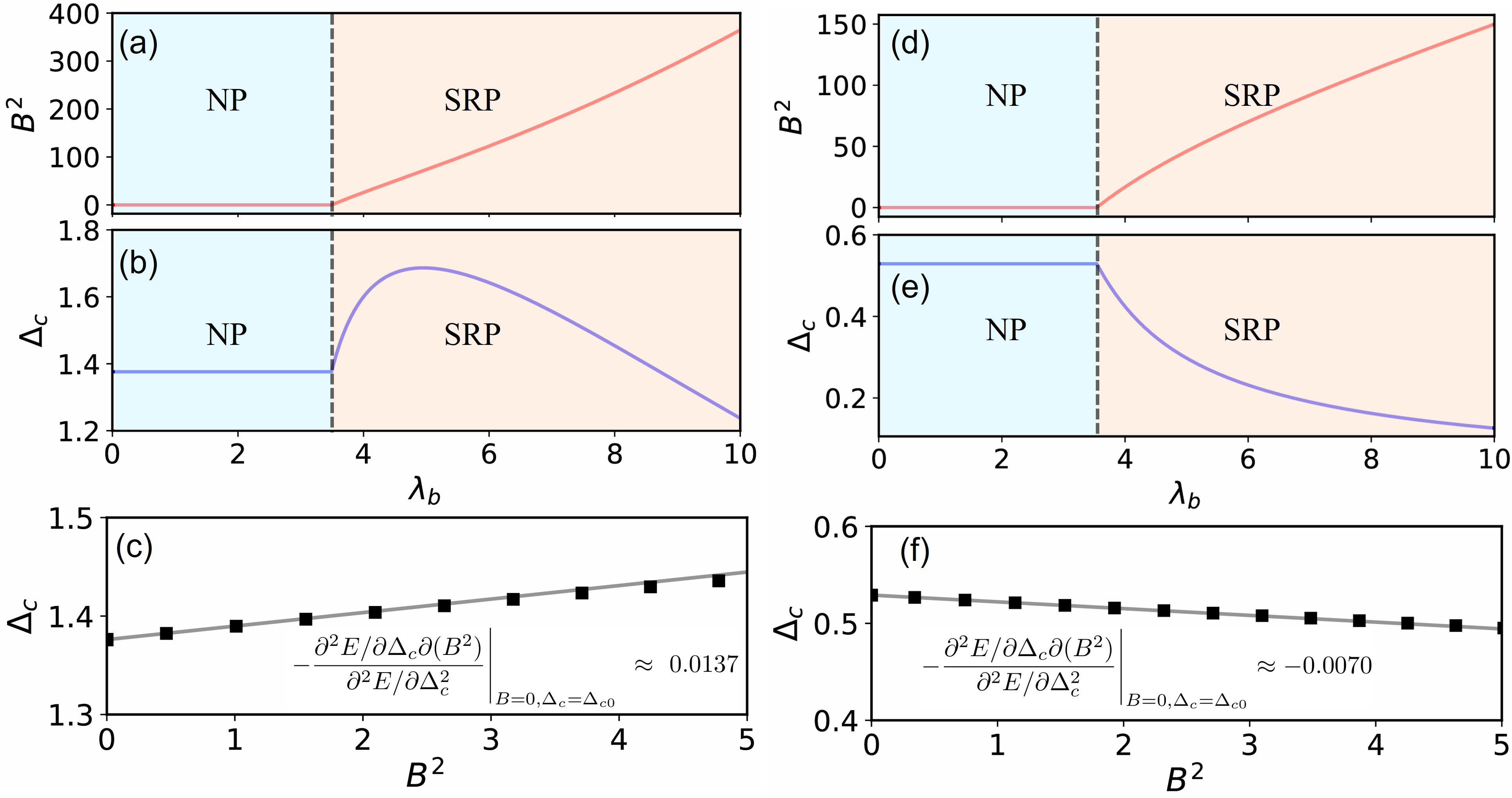}
	\caption{Here, we draw plots of the order parameters $B^2$ (a) and $\Delta_c$ (b) against the atom-light coupling strength $\lambda_b$ with fixed Kerr strength $\chi=\chi_3=5\times10^{-5}$, the critical behavior of which is given by (c). Also, we give the plots of $B^2$ (d) and $\Delta_c$ (e)with $\chi=\chi_3=2\times10^{-3}$. The black dash lines stand for the critcial points seperate the NP ($B=0, \Delta_c\neq0$) and the SRP ($B\neq0, \Delta_c\neq0$), which are colored following the same rule as before. The critical behavior is shown in (f). Here, we fixed $\tilde{\omega}_b=1,\tilde{\omega}_c=0.01,\Omega=100,\lambda_c=0.1$ to guarantee the hierarchy relations of classical oscillator limit.}\label{main_2mode_QRM_fig}
\end{figure}

\textit{Floquet engineered two-mode Rabi model.} As a simple toy model for fermionic pairing, we design a two-mode Rabi model where two spin 1/2 systems couple with two optical modes. The effective Hamiltonian gives as
\begin{align}\label{H_eff}
	&H_{eff}=\tilde{\omega}_bb^\dagger b+\tilde{\omega}_cc^\dagger c+\dfrac{\Omega}{2}(\sigma_1^z+\sigma_2^z)\notag\\
	&+\lambda_b(b+b^\dagger)(\sigma_1^x+\sigma_2^x)-\dfrac{\lambda_c}{2}(c+c^\dagger)(\sigma_1^x+\sigma_2^x)^2.
\end{align}
$\tilde{\omega}_{b(c)}\equiv\omega_{b(c)}-\omega_{p}$ is the detuning between the cavity frequency $\omega_{b(c)}$ of the cavity eigenmode $b(c)$ and the pumping frequency $\omega_{p}$. Apart from the atom-light coupling term $\lambda_b(b+b^\dagger)(\sigma_1^x+\sigma_2^x)$, a higher order coupling term $-\dfrac{\lambda_c}{2}(c+c^\dagger)(\sigma_1^x+\sigma_2^x)^2$ is also taken into consideration, which can be designed by Floquet engineering~\cite{PhysRevLett.132.113402,PRXQuantum.4.020329,PhysRevLett.124.190601} as is shown in~\cite{supplementary} or the three-mode spontaneous parametric down-conversion in circuit QED systems~\cite{PhysRevX.10.011011,PhysRevApplied.10.044019,PhysRevLett.114.090503}. $\lambda_b$ and $\lambda_c$ are the effective coupling strengths of these two types of interactions.

In order to calculate the ground state energy and evaluate the potential phase transition, we apply the mean-field approximation which is valid in classic oscillator limit, where the hierarchy relation $\tilde{\omega}_{b(c)}\ll\lambda_{b(c)}\ll\Omega$ is required~\cite{PhysRevLett.115.180404,PhysRevA.101.063843}. Thus, the operator $b(c)$ is replaced into its expectation value as $b(c)\rightarrow\left<b(c)\right>$ and ignore the quantum fluctuation terms. Defining $\left<b(c)\right>\equiv B(C)$, it gives the ground state energy of the system as 
\begin{align}\label{E(B,C)}
	E(B,C)&=\tilde{\omega}_bB^2+\tilde{\omega}_cC^2-2\lambda_cC+\left<H_{spin}\right>+H_{\mathcal{K}}.
\end{align}
Here, we define the spin part Hamiltonian as $H_{spin}(B,C)\equiv\dfrac{\Omega}{2}(\sigma_1^z+\sigma_2^z)+2\lambda_bB(\sigma_1^x+\sigma_2^x)-2\lambda_cC\sigma_1^x\sigma_2^x$, and the last term will dominate the two pairing interaction, leading to the nonzero matrix element $\bra{\downarrow,\downarrow}H_{spin}\ket{\uparrow,\uparrow}$. Additionally, we consider the rather weak Kerr terms $H_{\mathcal{K}}=\chi_3B^2C^2+\chi(B^4+C^4)$ in our system~\cite{PhysRevLett.133.233604,RN92,PhysRevA.84.012329,PhysRevLett.132.053601}, thus we can obtain the critical boundary according to $0=\tilde{\omega}_b-\dfrac{8\lambda_b^2}{\Omega^2}(\sqrt{\Omega^2+\Delta_c^2}+\dfrac{\Delta_c^2}{\sqrt{\Omega^2+\Delta_c^2}}+2\Delta_c)+\dfrac{\chi_3\Delta_{c}^2}{4\lambda_c^2}$, where the pairing order is defined as $\Delta_c\equiv2\lambda_cC$. Following Eq.~(\ref{interplay}), the numerator and denominator are analytically accessible, given as
\begin{widetext}
	\begin{align}
		\label{partial_12}
		\dfrac{\partial^2E}{\partial\Delta_c\partial(B^2)}\Bigg|_{B=0,\Delta_c=\Delta_{c0}}&=\dfrac{\chi_3\Delta_{c0}}{2\lambda_c^2}-\dfrac{8\lambda_b^2}{\Omega^2}\left[2-\left(\dfrac{\Delta_{c0}}{\sqrt{\Omega^2+\Delta_{c0}^2}}\right)^3+\dfrac{3\Delta_{c0}}{\sqrt{\Omega^2+\Delta_{c0}^2}}\right]\\
		\label{partial_11}
		\dfrac{\partial^2E}{\partial\Delta_c^2}\Bigg|_{B=0,\Delta_c=\Delta_{c0}}&=\dfrac{\tilde{\omega}_c}{2\lambda_c^2}-\dfrac{1}{\sqrt{\Omega^2+\Delta_{c0}^2}}+\dfrac{\Delta_{c0}^2}{(\Omega^2+\Delta_{c0}^2)^{3/2}}+\dfrac{3\chi\Delta_{c0}^2}{4\lambda_c^4}.
	\end{align}
\end{widetext}
Thus we can obtain the linear fitting for the critical behavior in superradiant phase (SRP), reads as
\begin{align}\label{scaling_equation}
	\Delta_c=\Delta_{c0}-\dfrac{\partial^2E/\partial\Delta_c\partial(B^2)}{\partial^2E/\partial\Delta_c^2}\Bigg|_{B=0,\Delta_c=\Delta_{c0}}\times B^2,
\end{align}
with $\Delta_{c0}$ the pairing order in NP. It's found that larger $\chi_3$ tends to suppress the order parameter $\Delta_c$ by inserting Eqs.~(\ref{partial_12})(\ref{partial_11}) into Eq.~(\ref{scaling_equation}). 

The numerical results are shown in Fig.~\ref{main_2mode_QRM_fig}, which matches well with the linear fitting from the analytical results. The subfigure (a)(d) and (b)(e) show the change of the order parameter $B^2$ and $\Delta_c$ with different Kerr strength. In subfigure (b) the pairing strength $\Delta_c$ will enhance with the superradiance emerging, and then gradually decrease around $\lambda_b\approx5$. The scaling behavior is well predicted by the Eq.~\ref{scaling_equation} around the critcial point, shown in (c) with $\dfrac{\partial^2E}{\partial\Delta_c\partial(B^2)}\Bigg|_{B=0,\Delta_c=\Delta_{c0}}\approx-0.0165$ and $-\dfrac{\partial^2E/\partial\Delta_c\partial(B^2)}{\partial^2E/\partial\Delta_c^2}\Bigg|_{B=0,\Delta_c=\Delta_{c0}}\approx0.0137$. As the superradiant order $B$ increases with $\lambda_b$, the Kerr effect $\chi_3B^2C^2$ will further suppress the pairing order $C$. With $\lambda_b\gtrsim5$, such suppression will dominate the evolution of $\Delta_c$, leading to the decreasing eventually. In subfigure (e), the pairing order parameter will decrease from the beginning of the superradiance, whose scaling rate also matches well with the analytical results as is shown in (f) with $\dfrac{\partial^2E}{\partial\Delta_c\partial(B^2)}\Bigg|_{B=0,\Delta_c=\Delta_{c0}}\approx0.0327$ and $-\dfrac{\partial^2E/\partial\Delta_c\partial(B^2)}{\partial^2E/\partial\Delta_c^2}\Bigg|_{B=0,\Delta_c=\Delta_{c0}}\approx-0.0070$.

\textit{The 1D attractive Fermi Dicke model.}
For a more realistic model, we consider both the attractive interaction and the atom-light coupling in a fermionic ensemble, which can be simulated in optical lattice~\cite{RN103,doi:10.1126/science.abd4385}. The Hamiltonian consists of the single particle part $\tilde{H}_{single}$ and the interaction part $H_{FF}$, $H_{tot}=\tilde{H}_{single}+H_{FF}$~\cite{supplementary}, which reads
\begin{align}
	&\tilde{H}_{single}=\tilde{\omega}a^\dagger a+\sum_{\sigma,k}\dfrac{k^2}{2m}c_{\sigma,k}^\dagger c_{\sigma,k}\notag\\
	&-[\dfrac{g\Omega}{4\delta\sqrt{2N}}(a+a^\dagger)\sum_{k,s=\pm1}c_{\uparrow,k}^\dagger c_{\downarrow,k+sk_c}+H.c.]\\
	&H_{FF}\approx -\Delta\sum_{k}c^\dagger_{\uparrow,k}c^\dagger_{\downarrow,-k}-\Delta^*\sum_{k}c_{\downarrow,-k}c_{\uparrow,k}-\dfrac{2N\Delta^2}{U}.
\end{align}
Here, we only consider a 1D fermionic exsemble for simplicity. In thermodynamic limit, where the number of fermionic atom $2N\rightarrow\infty$, the mean-field approximation can be safely applied. By replacing the operator $a$ into $\left<a\right>\equiv\alpha$, the free energy operator can write as
\begin{align}
	\hat{F}&=\tilde{\omega}\left|\alpha\right|^2+\sum_{\sigma,k}(\dfrac{k^2}{2m}-\mu)c_{\sigma,k}^\dagger c_{\sigma,k}\notag\\
	&-[\dfrac{g\Omega}{4\delta\sqrt{2N}}(\alpha+\alpha^*)\sum_{k,s=\pm1}c_{\uparrow,k}^\dagger c_{\downarrow,k+sk_c}+H.c.]\notag\\
	&+\Delta\sum_{k}c^\dagger_{\uparrow,k}c^\dagger_{\downarrow,-k}+\Delta^*\sum_{k}c_{\downarrow,-k}c_{\uparrow,k}-\dfrac{2N\Delta^2}{U}+2\mu N.
\end{align}
Obviously, the two order parameters in this system are the superconducting order $\Delta$ and the superradiant order $\alpha$ respectively. The minimum point of the expectation value $F(\alpha,\Delta)=\left<\hat{F}(\alpha,\Delta)\right>$ in the parameter plane $(\alpha,\Delta)$ is acquired with $\dfrac{\partial F}{\partial\alpha}=\dfrac{\partial F}{\partial\Delta}=0$ satisfied. Because we only consider the non-dissipative case, it's easily checked that the minimum is always obtained with $\alpha=Re\alpha=\pm\left|\alpha\right|$. The chemical potential $\mu(\alpha,\Delta)$ is introduced to keep the total atom number conservative by $\dfrac{\partial F}{\partial\mu}=0$.

Consider the differentiation against the order parameter $\alpha^2$, the superradiant phase transition occurs at $\tilde{\omega}+4\chi=0$. $\chi$ is the susceptibility, which can be obtained by 2nd order pertubative theory as~\cite{supplementary}
\begin{widetext}
	\begin{align}
		\chi\equiv\dfrac{1}{4}\left(\dfrac{\partial F}{\partial\left|\alpha\right|^2}\Bigg|_{\alpha=0}-\tilde{\omega}\right)=-\left(\dfrac{g\Omega}{4\delta}\right)^2\dfrac{1}{2N}\sum_{k,s}\dfrac{v_k^2u_{-(k+sk_c)}^2+u_k^2v_{-(k+sk_c)}^2-2u_kv_ku_{-(k+sk_c)}v_{-(k+sk_c)}}{\sqrt{\left(\dfrac{k^2}{2m}-\mu\right)^2+\Delta^2}+\sqrt{\left(\dfrac{(k+sk_c)^2}{2m}-\mu\right)^2+\Delta^2}},
	\end{align}
\end{widetext}
with $u_k^2=\dfrac{1}{2}\left(1+\dfrac{k^2/2m-\mu}{\mathcal{E}_k}\right)$ and $v_k^2=\dfrac{1}{2}\left(1-\dfrac{k^2/2m-\mu}{\mathcal{E}_k}\right)$. And $\mathcal{E}_k=\sqrt{(\dfrac{k^2}{2m}-\mu)^2+\Delta^2}$ represents the excited energy of the Bogoliubov quasi-particle.

\begin{figure}[tb]
	\centering
	\hspace{-0.8cm}
	\includegraphics[width=0.52\textwidth]{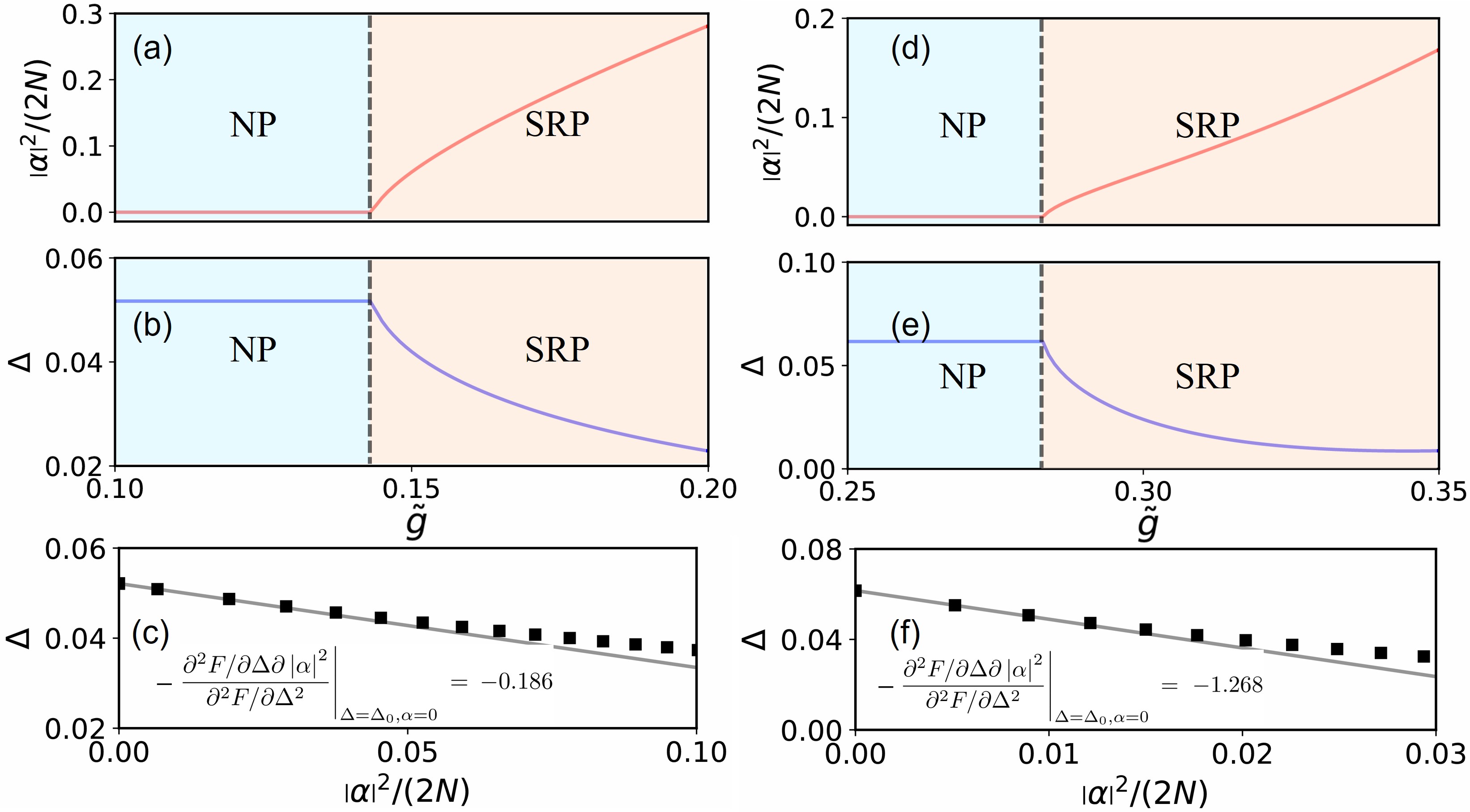}
	\caption{Here, we draw plots of the order parameters $\left|\alpha\right|^2$ (a) and $\Delta$ (b) changing against the dimensionless couping strength $\tilde{g}$ with small filling factor $k_F=0.1$ and the attractive potential $U=-0.05$. This gives us the pairing order $\Delta_0=0.0517$ and the critical coupling strength $\tilde{g}_0=0.143$; We also give the order parameters $\left|\alpha\right|^2$ (d) and $\Delta$ (e) with $k_F=0.8$ and the attractive potential $U=-0.6$, which gives us the pairing order $\Delta_0=0.0616$ and the critical coupling strength $\tilde{g}_c=0.283$. The black dash lines stand for the critcial points seperate the NP ($\alpha=0, \Delta_0\neq0$) and SRP ($\alpha\neq0, \Delta_0\neq0$), which are colored as the same rule before. The scaling rate is obtained by the 2nd order differentiation as -0.186 (c) and -1.268 (f) in these two cases respectively. In this example, we set $\tilde{\omega}=E_r$.}\label{main_1D_BCS}
\end{figure}

In Fig.~\ref{main_1D_BCS}, we give the numerical simulation of the order parameters $\alpha$ in subfigure (a)(d) and $\Delta$ in subfigure (b)(e) changing against the dimensionless atom-cavity coupling strength $\tilde{g}$ respectively, which is defined as $\tilde{g}\equiv\dfrac{\Omega^2g^2}{(4\delta)^2E_r^2}$ followed by the previous work~\cite{vph8-186g,PhysRevLett.112.143002,PhysRevLett.112.143003,PhysRevLett.112.143004}. Meanwhile, the recoil energy is set equal to the frequency of the boson mode as $\tilde{\omega}=E_r$ in the following discussion. Also, we fix the filling factor $k_F=0.1,~U=-0.05$ in (a)(b) and $k_F=0.8,~U=-0.6$ in (d)(e). It turns out that the superfluid order parameter $\Delta$ will increase with the emerging of superradiance as the coupling strength exceeding the 2nd order critical line (continous phase transition), marked by the black dashed lines. 

Although the denominate of the scaling rate $-\dfrac{\partial^2F/\partial\Delta\partial\left|\alpha\right|^2}{\partial^2F/\partial\Delta^2}\Bigg|_{\Delta=\Delta_0,\alpha=0}$ followed by Eq.~(\ref{interplay}) is hard to acquire, the calculation can also be accomplished by numerical finite difference. The numerator can be analytically obtained by $\partial^2F/\partial\Delta\partial\left|\alpha\right|^2\Bigg|_{\Delta=\Delta_0,\alpha=0}=4\dfrac{\partial\chi}{\partial\Delta}\Bigg|_{\Delta=\Delta_0}$, which tells us whether the pairing order will be suppressed or reinforced. As shown in (c)(f), the grey solid lines are the fitting lines with the ramping rates labeled on the figure.
The results match well with the black square points obtained by numerical simulation. Compare with the numerical optimization of the free energy with different dimensionless coupling strengths $\tilde{g}$, our scheme effectively saves the calculation resource and gives a scaling rate with rather high precision once obtaining the superfluid order parameter $\Delta_0$ in NP ($\alpha=0$) with arbitrary $\tilde{g}$.


Although the main idea of the letter is to give the universal critical behaviors around continuous phase boundary, this model also provides a good example where the superfluid order parameter jumps with the 1st order (discontinuous) superradiant phase transition, with the relevant discussion in~\cite{supplementary}.

Actually, the mean-field treatment of superconducting order in 1D system is insufficient to describe such system, beacuse the long-range order will be destroyed by strong quantum fluctuation according to Luttinger liquid theory~\cite{RN95,RN96,cui2026tomonagaluttingerliquidtheoryonedimensional}. But it still sheds light on the practicality and validity of our analytical results, let alone that the generalization to higher dimensional system is direct, where the mean-field approximation will be valid.

\textit{Summary.} In this work, we give the univeral critical behaviors of continuous phase transition in two order parameters case based
on Landau's theory, which is referred to the mean-field approximation. It serves as a good approximation in both classic oscillator limit and thermodynamic limit, although the quantum fluctuation is ignored. Through strict derivation, it's found that the value $-\dfrac{\partial_1\partial_2^{v_{m1}}F(\bar{O}_1^c,0;\vec{\lambda}_c)}{v_{m1}!\partial_1^2F(\bar{O}_1^c,0;\vec{\lambda}_c)}$ decides the changing rate of the order parameter $O_1$ right after the phase transition of $O_2$. It provides us a new idea to indrectly manipulate the required physical quantity by implementing phase transition of another order parameter. We introduce two examples in zero temperature limit to practice our scheme, where the strength of two-spin pairing and the superfluid order are modified by the occurrence of superradiance. However, we want to emphasize that our analytical scheme is only suitable for continuous critical boundary. Although the discontinuous phase transition provides us another avenue to realize the similar effect, the analytical prediction is absent.

Our work provides complete understanding and higher perspectives on complex systems described by two or more physcial quantities. It will benefit us to search for novel physical phenomena and explore new quantum materials through quantum manipulation and quantum simulation~\cite{PhysRevLett.84.4068,PhysRevA.104.053313,PhysRevLett.125.053602,PhysRevLett.127.177002,PhysRevLett.130.083603,PhysRevA.87.023831,PhysRevB.103.075131}, even in traditional condensed matter system.

\begin{acknowledgments}
	We thank Feng-xiao Sun for his useful suggestions for this work.
\end{acknowledgments}

\bibliography{ref}

\begin{thebibliography}{53}%
\makeatletter
\providecommand \@ifxundefined [1]{%
 \@ifx{#1\undefined}
}%
\providecommand \@ifnum [1]{%
 \ifnum #1\expandafter \@firstoftwo
 \else \expandafter \@secondoftwo
 \fi
}%
\providecommand \@ifx [1]{%
 \ifx #1\expandafter \@firstoftwo
 \else \expandafter \@secondoftwo
 \fi
}%
\providecommand \natexlab [1]{#1}%
\providecommand \enquote  [1]{``#1''}%
\providecommand \bibnamefont  [1]{#1}%
\providecommand \bibfnamefont [1]{#1}%
\providecommand \citenamefont [1]{#1}%
\providecommand \href@noop [0]{\@secondoftwo}%
\providecommand \href [0]{\begingroup \@sanitize@url \@href}%
\providecommand \@href[1]{\@@startlink{#1}\@@href}%
\providecommand \@@href[1]{\endgroup#1\@@endlink}%
\providecommand \@sanitize@url [0]{\catcode `\\12\catcode `\$12\catcode
  `\&12\catcode `\#12\catcode `\^12\catcode `\_12\catcode `\%12\relax}%
\providecommand \@@startlink[1]{}%
\providecommand \@@endlink[0]{}%
\providecommand \url  [0]{\begingroup\@sanitize@url \@url }%
\providecommand \@url [1]{\endgroup\@href {#1}{\urlprefix }}%
\providecommand \urlprefix  [0]{URL }%
\providecommand \Eprint [0]{\href }%
\providecommand \doibase [0]{http://dx.doi.org/}%
\providecommand \selectlanguage [0]{\@gobble}%
\providecommand \bibinfo  [0]{\@secondoftwo}%
\providecommand \bibfield  [0]{\@secondoftwo}%
\providecommand \translation [1]{[#1]}%
\providecommand \BibitemOpen [0]{}%
\providecommand \bibitemStop [0]{}%
\providecommand \bibitemNoStop [0]{.\EOS\space}%
\providecommand \EOS [0]{\spacefactor3000\relax}%
\providecommand \BibitemShut  [1]{\csname bibitem#1\endcsname}%
\let\auto@bib@innerbib\@empty
\bibitem [{\citenamefont {Peierls}(1936)}]{Peierls1936}%
  \BibitemOpen
  \bibfield  {author} {\bibinfo {author} {\bibfnamefont {R.}~\bibnamefont
  {Peierls}},\ }\href {\doibase 10.1017/S0305004100019174} {\bibfield
  {journal} {\bibinfo  {journal} {Mathematical Proceedings of the Cambridge
  Philosophical Society}\ }\textbf {\bibinfo {volume} {32}},\ \bibinfo {pages}
  {477–481} (\bibinfo {year} {1936})}\BibitemShut {NoStop}%
\bibitem [{\citenamefont {Stanley}(1968)}]{PhysRevLett.20.589}%
  \BibitemOpen
  \bibfield  {author} {\bibinfo {author} {\bibfnamefont {H.~E.}\ \bibnamefont
  {Stanley}},\ }\href {\doibase 10.1103/PhysRevLett.20.589} {\bibfield
  {journal} {\bibinfo  {journal} {Phys. Rev. Lett.}\ }\textbf {\bibinfo
  {volume} {20}},\ \bibinfo {pages} {589} (\bibinfo {year} {1968})}\BibitemShut
  {NoStop}%
\bibitem [{\citenamefont {Aizenman}\ and\ \citenamefont
  {Simon}(1980)}]{AIZENMAN1980281}%
  \BibitemOpen
  \bibfield  {author} {\bibinfo {author} {\bibfnamefont {M.}~\bibnamefont
  {Aizenman}}\ and\ \bibinfo {author} {\bibfnamefont {B.}~\bibnamefont
  {Simon}},\ }\href {\doibase https://doi.org/10.1016/0375-9601(80)90493-4}
  {\bibfield  {journal} {\bibinfo  {journal} {Physics Letters A}\ }\textbf
  {\bibinfo {volume} {76}},\ \bibinfo {pages} {281} (\bibinfo {year}
  {1980})}\BibitemShut {NoStop}%
\bibitem [{\citenamefont {Hohenberg}\ and\ \citenamefont
  {Krekhov}(2015)}]{HOHENBERG20151}%
  \BibitemOpen
  \bibfield  {author} {\bibinfo {author} {\bibfnamefont {P.}~\bibnamefont
  {Hohenberg}}\ and\ \bibinfo {author} {\bibfnamefont {A.}~\bibnamefont
  {Krekhov}},\ }\href {\doibase https://doi.org/10.1016/j.physrep.2015.01.001}
  {\bibfield  {journal} {\bibinfo  {journal} {Physics Reports}\ }\textbf
  {\bibinfo {volume} {572}},\ \bibinfo {pages} {1} (\bibinfo {year} {2015})},\
  \bibinfo {note} {an introduction to the Ginzburg–Landau theory of phase
  transitions and nonequilibrium patterns}\BibitemShut {NoStop}%
\bibitem [{\citenamefont {Hwang}\ \emph {et~al.}(2015)\citenamefont {Hwang},
  \citenamefont {Puebla},\ and\ \citenamefont
  {Plenio}}]{PhysRevLett.115.180404}%
  \BibitemOpen
  \bibfield  {author} {\bibinfo {author} {\bibfnamefont {M.-J.}\ \bibnamefont
  {Hwang}}, \bibinfo {author} {\bibfnamefont {R.}~\bibnamefont {Puebla}}, \
  and\ \bibinfo {author} {\bibfnamefont {M.~B.}\ \bibnamefont {Plenio}},\
  }\href {\doibase 10.1103/PhysRevLett.115.180404} {\bibfield  {journal}
  {\bibinfo  {journal} {Phys. Rev. Lett.}\ }\textbf {\bibinfo {volume} {115}},\
  \bibinfo {pages} {180404} (\bibinfo {year} {2015})}\BibitemShut {NoStop}%
\bibitem [{\citenamefont {Hwang}\ and\ \citenamefont
  {Plenio}(2016)}]{PhysRevLett.117.123602}%
  \BibitemOpen
  \bibfield  {author} {\bibinfo {author} {\bibfnamefont {M.-J.}\ \bibnamefont
  {Hwang}}\ and\ \bibinfo {author} {\bibfnamefont {M.~B.}\ \bibnamefont
  {Plenio}},\ }\href {\doibase 10.1103/PhysRevLett.117.123602} {\bibfield
  {journal} {\bibinfo  {journal} {Phys. Rev. Lett.}\ }\textbf {\bibinfo
  {volume} {117}},\ \bibinfo {pages} {123602} (\bibinfo {year}
  {2016})}\BibitemShut {NoStop}%
\bibitem [{\citenamefont {Liu}\ \emph {et~al.}(2017)\citenamefont {Liu},
  \citenamefont {Chesi}, \citenamefont {Ying}, \citenamefont {Chen},
  \citenamefont {Luo},\ and\ \citenamefont {Lin}}]{PhysRevLett.119.220601}%
  \BibitemOpen
  \bibfield  {author} {\bibinfo {author} {\bibfnamefont {M.}~\bibnamefont
  {Liu}}, \bibinfo {author} {\bibfnamefont {S.}~\bibnamefont {Chesi}}, \bibinfo
  {author} {\bibfnamefont {Z.-J.}\ \bibnamefont {Ying}}, \bibinfo {author}
  {\bibfnamefont {X.}~\bibnamefont {Chen}}, \bibinfo {author} {\bibfnamefont
  {H.-G.}\ \bibnamefont {Luo}}, \ and\ \bibinfo {author} {\bibfnamefont
  {H.-Q.}\ \bibnamefont {Lin}},\ }\href {\doibase
  10.1103/PhysRevLett.119.220601} {\bibfield  {journal} {\bibinfo  {journal}
  {Phys. Rev. Lett.}\ }\textbf {\bibinfo {volume} {119}},\ \bibinfo {pages}
  {220601} (\bibinfo {year} {2017})}\BibitemShut {NoStop}%
\bibitem [{\citenamefont {Nagy}\ \emph {et~al.}(2010)\citenamefont {Nagy},
  \citenamefont {K\'onya}, \citenamefont {Szirmai},\ and\ \citenamefont
  {Domokos}}]{PhysRevLett.104.130401}%
  \BibitemOpen
  \bibfield  {author} {\bibinfo {author} {\bibfnamefont {D.}~\bibnamefont
  {Nagy}}, \bibinfo {author} {\bibfnamefont {G.}~\bibnamefont {K\'onya}},
  \bibinfo {author} {\bibfnamefont {G.}~\bibnamefont {Szirmai}}, \ and\
  \bibinfo {author} {\bibfnamefont {P.}~\bibnamefont {Domokos}},\ }\href
  {\doibase 10.1103/PhysRevLett.104.130401} {\bibfield  {journal} {\bibinfo
  {journal} {Phys. Rev. Lett.}\ }\textbf {\bibinfo {volume} {104}},\ \bibinfo
  {pages} {130401} (\bibinfo {year} {2010})}\BibitemShut {NoStop}%
\bibitem [{\citenamefont {Xu}\ and\ \citenamefont
  {Pu}(2019)}]{PhysRevLett.122.193201}%
  \BibitemOpen
  \bibfield  {author} {\bibinfo {author} {\bibfnamefont {Y.}~\bibnamefont
  {Xu}}\ and\ \bibinfo {author} {\bibfnamefont {H.}~\bibnamefont {Pu}},\ }\href
  {\doibase 10.1103/PhysRevLett.122.193201} {\bibfield  {journal} {\bibinfo
  {journal} {Phys. Rev. Lett.}\ }\textbf {\bibinfo {volume} {122}},\ \bibinfo
  {pages} {193201} (\bibinfo {year} {2019})}\BibitemShut {NoStop}%
\bibitem [{\citenamefont {Soriente}\ \emph {et~al.}(2018)\citenamefont
  {Soriente}, \citenamefont {Donner}, \citenamefont {Chitra},\ and\
  \citenamefont {Zilberberg}}]{PhysRevLett.120.183603}%
  \BibitemOpen
  \bibfield  {author} {\bibinfo {author} {\bibfnamefont {M.}~\bibnamefont
  {Soriente}}, \bibinfo {author} {\bibfnamefont {T.}~\bibnamefont {Donner}},
  \bibinfo {author} {\bibfnamefont {R.}~\bibnamefont {Chitra}}, \ and\ \bibinfo
  {author} {\bibfnamefont {O.}~\bibnamefont {Zilberberg}},\ }\href {\doibase
  10.1103/PhysRevLett.120.183603} {\bibfield  {journal} {\bibinfo  {journal}
  {Phys. Rev. Lett.}\ }\textbf {\bibinfo {volume} {120}},\ \bibinfo {pages}
  {183603} (\bibinfo {year} {2018})}\BibitemShut {NoStop}%
\bibitem [{\citenamefont {Baksic}\ and\ \citenamefont
  {Ciuti}(2014)}]{PhysRevLett.112.173601}%
  \BibitemOpen
  \bibfield  {author} {\bibinfo {author} {\bibfnamefont {A.}~\bibnamefont
  {Baksic}}\ and\ \bibinfo {author} {\bibfnamefont {C.}~\bibnamefont {Ciuti}},\
  }\href {\doibase 10.1103/PhysRevLett.112.173601} {\bibfield  {journal}
  {\bibinfo  {journal} {Phys. Rev. Lett.}\ }\textbf {\bibinfo {volume} {112}},\
  \bibinfo {pages} {173601} (\bibinfo {year} {2014})}\BibitemShut {NoStop}%
\bibitem [{\citenamefont {Bastidas}\ \emph {et~al.}(2012)\citenamefont
  {Bastidas}, \citenamefont {Emary}, \citenamefont {Regler},\ and\
  \citenamefont {Brandes}}]{PhysRevLett.108.043003}%
  \BibitemOpen
  \bibfield  {author} {\bibinfo {author} {\bibfnamefont {V.~M.}\ \bibnamefont
  {Bastidas}}, \bibinfo {author} {\bibfnamefont {C.}~\bibnamefont {Emary}},
  \bibinfo {author} {\bibfnamefont {B.}~\bibnamefont {Regler}}, \ and\ \bibinfo
  {author} {\bibfnamefont {T.}~\bibnamefont {Brandes}},\ }\href {\doibase
  10.1103/PhysRevLett.108.043003} {\bibfield  {journal} {\bibinfo  {journal}
  {Phys. Rev. Lett.}\ }\textbf {\bibinfo {volume} {108}},\ \bibinfo {pages}
  {043003} (\bibinfo {year} {2012})}\BibitemShut {NoStop}%
\bibitem [{\citenamefont {Zhu}\ \emph {et~al.}(2020)\citenamefont {Zhu},
  \citenamefont {Ping}, \citenamefont {Yang},\ and\ \citenamefont
  {Agarwal}}]{PhysRevLett.124.073602}%
  \BibitemOpen
  \bibfield  {author} {\bibinfo {author} {\bibfnamefont {C.~J.}\ \bibnamefont
  {Zhu}}, \bibinfo {author} {\bibfnamefont {L.~L.}\ \bibnamefont {Ping}},
  \bibinfo {author} {\bibfnamefont {Y.~P.}\ \bibnamefont {Yang}}, \ and\
  \bibinfo {author} {\bibfnamefont {G.~S.}\ \bibnamefont {Agarwal}},\ }\href
  {\doibase 10.1103/PhysRevLett.124.073602} {\bibfield  {journal} {\bibinfo
  {journal} {Phys. Rev. Lett.}\ }\textbf {\bibinfo {volume} {124}},\ \bibinfo
  {pages} {073602} (\bibinfo {year} {2020})}\BibitemShut {NoStop}%
\bibitem [{\citenamefont {Mivehvar}(2024)}]{PhysRevLett.132.073602}%
  \BibitemOpen
  \bibfield  {author} {\bibinfo {author} {\bibfnamefont {F.}~\bibnamefont
  {Mivehvar}},\ }\href {\doibase 10.1103/PhysRevLett.132.073602} {\bibfield
  {journal} {\bibinfo  {journal} {Phys. Rev. Lett.}\ }\textbf {\bibinfo
  {volume} {132}},\ \bibinfo {pages} {073602} (\bibinfo {year}
  {2024})}\BibitemShut {NoStop}%
\bibitem [{\citenamefont {Greiner}\ \emph {et~al.}(2002)\citenamefont
  {Greiner}, \citenamefont {Mandel}, \citenamefont {Esslinger}, \citenamefont
  {Hänsch},\ and\ \citenamefont {Bloch}}]{RN105}%
  \BibitemOpen
  \bibfield  {author} {\bibinfo {author} {\bibfnamefont {M.}~\bibnamefont
  {Greiner}}, \bibinfo {author} {\bibfnamefont {O.}~\bibnamefont {Mandel}},
  \bibinfo {author} {\bibfnamefont {T.}~\bibnamefont {Esslinger}}, \bibinfo
  {author} {\bibfnamefont {T.~W.}\ \bibnamefont {Hänsch}}, \ and\ \bibinfo
  {author} {\bibfnamefont {I.}~\bibnamefont {Bloch}},\ }\href {\doibase
  10.1038/415039a} {\bibfield  {journal} {\bibinfo  {journal} {Nature}\
  }\textbf {\bibinfo {volume} {415}},\ \bibinfo {pages} {39} (\bibinfo {year}
  {2002})}\BibitemShut {NoStop}%
\bibitem [{\citenamefont {Baumann}\ \emph {et~al.}(2010)\citenamefont
  {Baumann}, \citenamefont {Guerlin}, \citenamefont {Brennecke},\ and\
  \citenamefont {Esslinger}}]{RN103}%
  \BibitemOpen
  \bibfield  {author} {\bibinfo {author} {\bibfnamefont {K.}~\bibnamefont
  {Baumann}}, \bibinfo {author} {\bibfnamefont {C.}~\bibnamefont {Guerlin}},
  \bibinfo {author} {\bibfnamefont {F.}~\bibnamefont {Brennecke}}, \ and\
  \bibinfo {author} {\bibfnamefont {T.}~\bibnamefont {Esslinger}},\ }\href
  {\doibase 10.1038/nature09009} {\bibfield  {journal} {\bibinfo  {journal}
  {Nature}\ }\textbf {\bibinfo {volume} {464}},\ \bibinfo {pages} {1301}
  (\bibinfo {year} {2010})}\BibitemShut {NoStop}%
\bibitem [{\citenamefont {Gopalakrishnan}\ \emph {et~al.}(2009)\citenamefont
  {Gopalakrishnan}, \citenamefont {Lev},\ and\ \citenamefont
  {Goldbart}}]{RN104}%
  \BibitemOpen
  \bibfield  {author} {\bibinfo {author} {\bibfnamefont {S.}~\bibnamefont
  {Gopalakrishnan}}, \bibinfo {author} {\bibfnamefont {B.~L.}\ \bibnamefont
  {Lev}}, \ and\ \bibinfo {author} {\bibfnamefont {P.~M.}\ \bibnamefont
  {Goldbart}},\ }\href {\doibase 10.1038/nphys1403} {\bibfield  {journal}
  {\bibinfo  {journal} {Nat. Phys.}\ }\textbf {\bibinfo {volume} {5}},\
  \bibinfo {pages} {845} (\bibinfo {year} {2009})}\BibitemShut {NoStop}%
\bibitem [{\citenamefont {Keeling}\ \emph {et~al.}(2010)\citenamefont
  {Keeling}, \citenamefont {Bhaseen},\ and\ \citenamefont
  {Simons}}]{PhysRevLett.105.043001}%
  \BibitemOpen
  \bibfield  {author} {\bibinfo {author} {\bibfnamefont {J.}~\bibnamefont
  {Keeling}}, \bibinfo {author} {\bibfnamefont {M.~J.}\ \bibnamefont
  {Bhaseen}}, \ and\ \bibinfo {author} {\bibfnamefont {B.~D.}\ \bibnamefont
  {Simons}},\ }\href {\doibase 10.1103/PhysRevLett.105.043001} {\bibfield
  {journal} {\bibinfo  {journal} {Phys. Rev. Lett.}\ }\textbf {\bibinfo
  {volume} {105}},\ \bibinfo {pages} {043001} (\bibinfo {year}
  {2010})}\BibitemShut {NoStop}%
\bibitem [{\citenamefont {St\"oferle}\ \emph {et~al.}(2004)\citenamefont
  {St\"oferle}, \citenamefont {Moritz}, \citenamefont {Schori}, \citenamefont
  {K\"ohl},\ and\ \citenamefont {Esslinger}}]{PhysRevLett.92.130403}%
  \BibitemOpen
  \bibfield  {author} {\bibinfo {author} {\bibfnamefont {T.}~\bibnamefont
  {St\"oferle}}, \bibinfo {author} {\bibfnamefont {H.}~\bibnamefont {Moritz}},
  \bibinfo {author} {\bibfnamefont {C.}~\bibnamefont {Schori}}, \bibinfo
  {author} {\bibfnamefont {M.}~\bibnamefont {K\"ohl}}, \ and\ \bibinfo {author}
  {\bibfnamefont {T.}~\bibnamefont {Esslinger}},\ }\href {\doibase
  10.1103/PhysRevLett.92.130403} {\bibfield  {journal} {\bibinfo  {journal}
  {Phys. Rev. Lett.}\ }\textbf {\bibinfo {volume} {92}},\ \bibinfo {pages}
  {130403} (\bibinfo {year} {2004})}\BibitemShut {NoStop}%
\bibitem [{\citenamefont {Landig}\ \emph {et~al.}(2016)\citenamefont {Landig},
  \citenamefont {Hruby}, \citenamefont {Dogra}, \citenamefont {Landini},
  \citenamefont {Mottl}, \citenamefont {Donner},\ and\ \citenamefont
  {Esslinger}}]{RN107}%
  \BibitemOpen
  \bibfield  {author} {\bibinfo {author} {\bibfnamefont {R.}~\bibnamefont
  {Landig}}, \bibinfo {author} {\bibfnamefont {L.}~\bibnamefont {Hruby}},
  \bibinfo {author} {\bibfnamefont {N.}~\bibnamefont {Dogra}}, \bibinfo
  {author} {\bibfnamefont {M.}~\bibnamefont {Landini}}, \bibinfo {author}
  {\bibfnamefont {R.}~\bibnamefont {Mottl}}, \bibinfo {author} {\bibfnamefont
  {T.}~\bibnamefont {Donner}}, \ and\ \bibinfo {author} {\bibfnamefont
  {T.}~\bibnamefont {Esslinger}},\ }\href {\doibase 10.1038/nature17409}
  {\bibfield  {journal} {\bibinfo  {journal} {Nature}\ }\textbf {\bibinfo
  {volume} {532}},\ \bibinfo {pages} {476} (\bibinfo {year}
  {2016})}\BibitemShut {NoStop}%
\bibitem [{\citenamefont {Esslinger}(2010)}]{annurev}%
  \BibitemOpen
  \bibfield  {author} {\bibinfo {author} {\bibfnamefont {T.}~\bibnamefont
  {Esslinger}},\ }\href {\doibase
  https://doi.org/10.1146/annurev-conmatphys-070909-104059} {\bibfield
  {journal} {\bibinfo  {journal} {Annu. Rev. Condens. Matter Phys.}\ }\textbf
  {\bibinfo {volume} {1}},\ \bibinfo {pages} {129} (\bibinfo {year}
  {2010})}\BibitemShut {NoStop}%
\bibitem [{\citenamefont {Keeling}\ \emph {et~al.}(2014)\citenamefont
  {Keeling}, \citenamefont {Bhaseen},\ and\ \citenamefont
  {Simons}}]{PhysRevLett.112.143002}%
  \BibitemOpen
  \bibfield  {author} {\bibinfo {author} {\bibfnamefont {J.}~\bibnamefont
  {Keeling}}, \bibinfo {author} {\bibfnamefont {M.~J.}\ \bibnamefont
  {Bhaseen}}, \ and\ \bibinfo {author} {\bibfnamefont {B.~D.}\ \bibnamefont
  {Simons}},\ }\href {\doibase 10.1103/PhysRevLett.112.143002} {\bibfield
  {journal} {\bibinfo  {journal} {Phys. Rev. Lett.}\ }\textbf {\bibinfo
  {volume} {112}},\ \bibinfo {pages} {143002} (\bibinfo {year}
  {2014})}\BibitemShut {NoStop}%
\bibitem [{\citenamefont {Piazza}\ and\ \citenamefont
  {Strack}(2014)}]{PhysRevLett.112.143003}%
  \BibitemOpen
  \bibfield  {author} {\bibinfo {author} {\bibfnamefont {F.}~\bibnamefont
  {Piazza}}\ and\ \bibinfo {author} {\bibfnamefont {P.}~\bibnamefont
  {Strack}},\ }\href {\doibase 10.1103/PhysRevLett.112.143003} {\bibfield
  {journal} {\bibinfo  {journal} {Phys. Rev. Lett.}\ }\textbf {\bibinfo
  {volume} {112}},\ \bibinfo {pages} {143003} (\bibinfo {year}
  {2014})}\BibitemShut {NoStop}%
\bibitem [{\citenamefont {Chen}\ \emph {et~al.}(2014)\citenamefont {Chen},
  \citenamefont {Yu},\ and\ \citenamefont {Zhai}}]{PhysRevLett.112.143004}%
  \BibitemOpen
  \bibfield  {author} {\bibinfo {author} {\bibfnamefont {Y.}~\bibnamefont
  {Chen}}, \bibinfo {author} {\bibfnamefont {Z.}~\bibnamefont {Yu}}, \ and\
  \bibinfo {author} {\bibfnamefont {H.}~\bibnamefont {Zhai}},\ }\href {\doibase
  10.1103/PhysRevLett.112.143004} {\bibfield  {journal} {\bibinfo  {journal}
  {Phys. Rev. Lett.}\ }\textbf {\bibinfo {volume} {112}},\ \bibinfo {pages}
  {143004} (\bibinfo {year} {2014})}\BibitemShut {NoStop}%
\bibitem [{\citenamefont {Zhang}\ \emph {et~al.}(2021)\citenamefont {Zhang},
  \citenamefont {Chen}, \citenamefont {Wu}, \citenamefont {Wang}, \citenamefont
  {Fan}, \citenamefont {Deng},\ and\ \citenamefont
  {Wu}}]{doi:10.1126/science.abd4385}%
  \BibitemOpen
  \bibfield  {author} {\bibinfo {author} {\bibfnamefont {X.}~\bibnamefont
  {Zhang}}, \bibinfo {author} {\bibfnamefont {Y.}~\bibnamefont {Chen}},
  \bibinfo {author} {\bibfnamefont {Z.}~\bibnamefont {Wu}}, \bibinfo {author}
  {\bibfnamefont {J.}~\bibnamefont {Wang}}, \bibinfo {author} {\bibfnamefont
  {J.}~\bibnamefont {Fan}}, \bibinfo {author} {\bibfnamefont {S.}~\bibnamefont
  {Deng}}, \ and\ \bibinfo {author} {\bibfnamefont {H.}~\bibnamefont {Wu}},\
  }\href {\doibase 10.1126/science.abd4385} {\bibfield  {journal} {\bibinfo
  {journal} {Science}\ }\textbf {\bibinfo {volume} {373}},\ \bibinfo {pages}
  {1359} (\bibinfo {year} {2021})}\BibitemShut {NoStop}%
\bibitem [{\citenamefont {Wu}\ \emph {et~al.}(2024)\citenamefont {Wu},
  \citenamefont {Hu}, \citenamefont {Wang}, \citenamefont {Chen}, \citenamefont
  {Li}, \citenamefont {Zhao}, \citenamefont {L\"u},\ and\ \citenamefont
  {Peng}}]{PhysRevLett.133.173602}%
  \BibitemOpen
  \bibfield  {author} {\bibinfo {author} {\bibfnamefont {Z.}~\bibnamefont
  {Wu}}, \bibinfo {author} {\bibfnamefont {C.}~\bibnamefont {Hu}}, \bibinfo
  {author} {\bibfnamefont {T.}~\bibnamefont {Wang}}, \bibinfo {author}
  {\bibfnamefont {Y.}~\bibnamefont {Chen}}, \bibinfo {author} {\bibfnamefont
  {Y.}~\bibnamefont {Li}}, \bibinfo {author} {\bibfnamefont {L.}~\bibnamefont
  {Zhao}}, \bibinfo {author} {\bibfnamefont {X.-Y.}\ \bibnamefont {L\"u}}, \
  and\ \bibinfo {author} {\bibfnamefont {X.}~\bibnamefont {Peng}},\ }\href
  {\doibase 10.1103/PhysRevLett.133.173602} {\bibfield  {journal} {\bibinfo
  {journal} {Phys. Rev. Lett.}\ }\textbf {\bibinfo {volume} {133}},\ \bibinfo
  {pages} {173602} (\bibinfo {year} {2024})}\BibitemShut {NoStop}%
\bibitem [{\citenamefont {Wang}\ and\ \citenamefont
  {Hioe}(1973)}]{PhysRevA.7.831}%
  \BibitemOpen
  \bibfield  {author} {\bibinfo {author} {\bibfnamefont {Y.~K.}\ \bibnamefont
  {Wang}}\ and\ \bibinfo {author} {\bibfnamefont {F.~T.}\ \bibnamefont
  {Hioe}},\ }\href {\doibase 10.1103/PhysRevA.7.831} {\bibfield  {journal}
  {\bibinfo  {journal} {Phys. Rev. A}\ }\textbf {\bibinfo {volume} {7}},\
  \bibinfo {pages} {831} (\bibinfo {year} {1973})}\BibitemShut {NoStop}%
\bibitem [{\citenamefont {Emary}\ and\ \citenamefont
  {Brandes}(2003)}]{PhysRevE.67.066203}%
  \BibitemOpen
  \bibfield  {author} {\bibinfo {author} {\bibfnamefont {C.}~\bibnamefont
  {Emary}}\ and\ \bibinfo {author} {\bibfnamefont {T.}~\bibnamefont
  {Brandes}},\ }\href {\doibase 10.1103/PhysRevE.67.066203} {\bibfield
  {journal} {\bibinfo  {journal} {Phys. Rev. E}\ }\textbf {\bibinfo {volume}
  {67}},\ \bibinfo {pages} {066203} (\bibinfo {year} {2003})}\BibitemShut
  {NoStop}%
\bibitem [{\citenamefont {Hepp}\ and\ \citenamefont
  {Lieb}(1973)}]{HEPP1973360}%
  \BibitemOpen
  \bibfield  {author} {\bibinfo {author} {\bibfnamefont {K.}~\bibnamefont
  {Hepp}}\ and\ \bibinfo {author} {\bibfnamefont {E.~H.}\ \bibnamefont
  {Lieb}},\ }\href {\doibase https://doi.org/10.1016/0003-4916(73)90039-0}
  {\bibfield  {journal} {\bibinfo  {journal} {Ann. Phys.}\ }\textbf {\bibinfo
  {volume} {76}},\ \bibinfo {pages} {360} (\bibinfo {year} {1973})}\BibitemShut
  {NoStop}%
\bibitem [{\citenamefont {Bardeen}\ \emph {et~al.}(1957)\citenamefont
  {Bardeen}, \citenamefont {Cooper},\ and\ \citenamefont
  {Schrieffer}}]{PhysRev.108.1175}%
  \BibitemOpen
  \bibfield  {author} {\bibinfo {author} {\bibfnamefont {J.}~\bibnamefont
  {Bardeen}}, \bibinfo {author} {\bibfnamefont {L.~N.}\ \bibnamefont {Cooper}},
  \ and\ \bibinfo {author} {\bibfnamefont {J.~R.}\ \bibnamefont {Schrieffer}},\
  }\href {\doibase 10.1103/PhysRev.108.1175} {\bibfield  {journal} {\bibinfo
  {journal} {Phys. Rev.}\ }\textbf {\bibinfo {volume} {108}},\ \bibinfo {pages}
  {1175} (\bibinfo {year} {1957})}\BibitemShut {NoStop}%
\bibitem [{\citenamefont {Zhang}\ \emph {et~al.}(2024)\citenamefont {Zhang},
  \citenamefont {Hu},\ and\ \citenamefont {Liu}}]{PhysRevLett.132.113402}%
  \BibitemOpen
  \bibfield  {author} {\bibinfo {author} {\bibfnamefont {X.}~\bibnamefont
  {Zhang}}, \bibinfo {author} {\bibfnamefont {Z.}~\bibnamefont {Hu}}, \ and\
  \bibinfo {author} {\bibfnamefont {Y.-C.}\ \bibnamefont {Liu}},\ }\href
  {\doibase 10.1103/PhysRevLett.132.113402} {\bibfield  {journal} {\bibinfo
  {journal} {Phys. Rev. Lett.}\ }\textbf {\bibinfo {volume} {132}},\ \bibinfo
  {pages} {113402} (\bibinfo {year} {2024})}\BibitemShut {NoStop}%
\bibitem [{\citenamefont {Sun}\ \emph {et~al.}(2023)\citenamefont {Sun},
  \citenamefont {Goldman}, \citenamefont {Aidelsburger},\ and\ \citenamefont
  {Bukov}}]{PRXQuantum.4.020329}%
  \BibitemOpen
  \bibfield  {author} {\bibinfo {author} {\bibfnamefont {B.-Y.}\ \bibnamefont
  {Sun}}, \bibinfo {author} {\bibfnamefont {N.}~\bibnamefont {Goldman}},
  \bibinfo {author} {\bibfnamefont {M.}~\bibnamefont {Aidelsburger}}, \ and\
  \bibinfo {author} {\bibfnamefont {M.}~\bibnamefont {Bukov}},\ }\href
  {\doibase 10.1103/PRXQuantum.4.020329} {\bibfield  {journal} {\bibinfo
  {journal} {PRX Quantum}\ }\textbf {\bibinfo {volume} {4}},\ \bibinfo {pages}
  {020329} (\bibinfo {year} {2023})}\BibitemShut {NoStop}%
\bibitem [{\citenamefont {Decker}\ \emph {et~al.}(2020)\citenamefont {Decker},
  \citenamefont {Karrasch}, \citenamefont {Eisert},\ and\ \citenamefont
  {Kennes}}]{PhysRevLett.124.190601}%
  \BibitemOpen
  \bibfield  {author} {\bibinfo {author} {\bibfnamefont {K.~S.~C.}\
  \bibnamefont {Decker}}, \bibinfo {author} {\bibfnamefont {C.}~\bibnamefont
  {Karrasch}}, \bibinfo {author} {\bibfnamefont {J.}~\bibnamefont {Eisert}}, \
  and\ \bibinfo {author} {\bibfnamefont {D.~M.}\ \bibnamefont {Kennes}},\
  }\href {\doibase 10.1103/PhysRevLett.124.190601} {\bibfield  {journal}
  {\bibinfo  {journal} {Phys. Rev. Lett.}\ }\textbf {\bibinfo {volume} {124}},\
  \bibinfo {pages} {190601} (\bibinfo {year} {2020})}\BibitemShut {NoStop}%
\bibitem [{sup()}]{supplementary}%
  \BibitemOpen
  \href@noop {} {}\bibinfo {note} {See Supplemental Material for further
  information on the theoretical treatment on the effective Hamiltonian of the
  two models and the discussion of the 1st order phase transition.}\BibitemShut
  {Stop}%
\bibitem [{\citenamefont {Chang}\ \emph {et~al.}(2020)\citenamefont {Chang},
  \citenamefont {Sab\'{\i}n}, \citenamefont {Forn-D\'{\i}az}, \citenamefont
  {Quijandr\'{\i}a}, \citenamefont {Vadiraj}, \citenamefont {Nsanzineza},
  \citenamefont {Johansson},\ and\ \citenamefont
  {Wilson}}]{PhysRevX.10.011011}%
  \BibitemOpen
  \bibfield  {author} {\bibinfo {author} {\bibfnamefont {C.~W.~S.}\
  \bibnamefont {Chang}}, \bibinfo {author} {\bibfnamefont {C.}~\bibnamefont
  {Sab\'{\i}n}}, \bibinfo {author} {\bibfnamefont {P.}~\bibnamefont
  {Forn-D\'{\i}az}}, \bibinfo {author} {\bibfnamefont {F.}~\bibnamefont
  {Quijandr\'{\i}a}}, \bibinfo {author} {\bibfnamefont {A.~M.}\ \bibnamefont
  {Vadiraj}}, \bibinfo {author} {\bibfnamefont {I.}~\bibnamefont {Nsanzineza}},
  \bibinfo {author} {\bibfnamefont {G.}~\bibnamefont {Johansson}}, \ and\
  \bibinfo {author} {\bibfnamefont {C.~M.}\ \bibnamefont {Wilson}},\ }\href
  {\doibase 10.1103/PhysRevX.10.011011} {\bibfield  {journal} {\bibinfo
  {journal} {Phys. Rev. X}\ }\textbf {\bibinfo {volume} {10}},\ \bibinfo
  {pages} {011011} (\bibinfo {year} {2020})}\BibitemShut {NoStop}%
\bibitem [{\citenamefont {Sandbo~Chang}\ \emph {et~al.}(2018)\citenamefont
  {Sandbo~Chang}, \citenamefont {Simoen}, \citenamefont {Aumentado},
  \citenamefont {Sab\'{\i}n}, \citenamefont {Forn-D\'{\i}az}, \citenamefont
  {Vadiraj}, \citenamefont {Quijandr\'{\i}a}, \citenamefont {Johansson},
  \citenamefont {Fuentes},\ and\ \citenamefont
  {Wilson}}]{PhysRevApplied.10.044019}%
  \BibitemOpen
  \bibfield  {author} {\bibinfo {author} {\bibfnamefont {C.~W.}\ \bibnamefont
  {Sandbo~Chang}}, \bibinfo {author} {\bibfnamefont {M.}~\bibnamefont
  {Simoen}}, \bibinfo {author} {\bibfnamefont {J.}~\bibnamefont {Aumentado}},
  \bibinfo {author} {\bibfnamefont {C.}~\bibnamefont {Sab\'{\i}n}}, \bibinfo
  {author} {\bibfnamefont {P.}~\bibnamefont {Forn-D\'{\i}az}}, \bibinfo
  {author} {\bibfnamefont {A.~M.}\ \bibnamefont {Vadiraj}}, \bibinfo {author}
  {\bibfnamefont {F.}~\bibnamefont {Quijandr\'{\i}a}}, \bibinfo {author}
  {\bibfnamefont {G.}~\bibnamefont {Johansson}}, \bibinfo {author}
  {\bibfnamefont {I.}~\bibnamefont {Fuentes}}, \ and\ \bibinfo {author}
  {\bibfnamefont {C.~M.}\ \bibnamefont {Wilson}},\ }\href {\doibase
  10.1103/PhysRevApplied.10.044019} {\bibfield  {journal} {\bibinfo  {journal}
  {Phys. Rev. Appl.}\ }\textbf {\bibinfo {volume} {10}},\ \bibinfo {pages}
  {044019} (\bibinfo {year} {2018})}\BibitemShut {NoStop}%
\bibitem [{\citenamefont {Flurin}\ \emph {et~al.}(2015)\citenamefont {Flurin},
  \citenamefont {Roch}, \citenamefont {Pillet}, \citenamefont {Mallet},\ and\
  \citenamefont {Huard}}]{PhysRevLett.114.090503}%
  \BibitemOpen
  \bibfield  {author} {\bibinfo {author} {\bibfnamefont {E.}~\bibnamefont
  {Flurin}}, \bibinfo {author} {\bibfnamefont {N.}~\bibnamefont {Roch}},
  \bibinfo {author} {\bibfnamefont {J.~D.}\ \bibnamefont {Pillet}}, \bibinfo
  {author} {\bibfnamefont {F.}~\bibnamefont {Mallet}}, \ and\ \bibinfo {author}
  {\bibfnamefont {B.}~\bibnamefont {Huard}},\ }\href {\doibase
  10.1103/PhysRevLett.114.090503} {\bibfield  {journal} {\bibinfo  {journal}
  {Phys. Rev. Lett.}\ }\textbf {\bibinfo {volume} {114}},\ \bibinfo {pages}
  {090503} (\bibinfo {year} {2015})}\BibitemShut {NoStop}%
\bibitem [{\citenamefont {Wang}\ \emph {et~al.}(2020)\citenamefont {Wang},
  \citenamefont {Liu}, \citenamefont {You}, \citenamefont {Chesi},
  \citenamefont {Luo},\ and\ \citenamefont {Lin}}]{PhysRevA.101.063843}%
  \BibitemOpen
  \bibfield  {author} {\bibinfo {author} {\bibfnamefont {Y.}~\bibnamefont
  {Wang}}, \bibinfo {author} {\bibfnamefont {M.}~\bibnamefont {Liu}}, \bibinfo
  {author} {\bibfnamefont {W.-L.}\ \bibnamefont {You}}, \bibinfo {author}
  {\bibfnamefont {S.}~\bibnamefont {Chesi}}, \bibinfo {author} {\bibfnamefont
  {H.-G.}\ \bibnamefont {Luo}}, \ and\ \bibinfo {author} {\bibfnamefont
  {H.-Q.}\ \bibnamefont {Lin}},\ }\href {\doibase 10.1103/PhysRevA.101.063843}
  {\bibfield  {journal} {\bibinfo  {journal} {Phys. Rev. A}\ }\textbf {\bibinfo
  {volume} {101}},\ \bibinfo {pages} {063843} (\bibinfo {year}
  {2020})}\BibitemShut {NoStop}%
\bibitem [{\citenamefont {Xu}\ \emph {et~al.}(2024)\citenamefont {Xu},
  \citenamefont {Sun}, \citenamefont {Zhang}, \citenamefont {He},\ and\
  \citenamefont {Pu}}]{PhysRevLett.133.233604}%
  \BibitemOpen
  \bibfield  {author} {\bibinfo {author} {\bibfnamefont {Y.}~\bibnamefont
  {Xu}}, \bibinfo {author} {\bibfnamefont {F.-X.}\ \bibnamefont {Sun}},
  \bibinfo {author} {\bibfnamefont {W.}~\bibnamefont {Zhang}}, \bibinfo
  {author} {\bibfnamefont {Q.}~\bibnamefont {He}}, \ and\ \bibinfo {author}
  {\bibfnamefont {H.}~\bibnamefont {Pu}},\ }\href {\doibase
  10.1103/PhysRevLett.133.233604} {\bibfield  {journal} {\bibinfo  {journal}
  {Phys. Rev. Lett.}\ }\textbf {\bibinfo {volume} {133}},\ \bibinfo {pages}
  {233604} (\bibinfo {year} {2024})}\BibitemShut {NoStop}%
\bibitem [{\citenamefont {Kounalakis}\ \emph {et~al.}(2018)\citenamefont
  {Kounalakis}, \citenamefont {Dickel}, \citenamefont {Bruno}, \citenamefont
  {Langford},\ and\ \citenamefont {Steele}}]{RN92}%
  \BibitemOpen
  \bibfield  {author} {\bibinfo {author} {\bibfnamefont {M.}~\bibnamefont
  {Kounalakis}}, \bibinfo {author} {\bibfnamefont {C.}~\bibnamefont {Dickel}},
  \bibinfo {author} {\bibfnamefont {A.}~\bibnamefont {Bruno}}, \bibinfo
  {author} {\bibfnamefont {N.~K.}\ \bibnamefont {Langford}}, \ and\ \bibinfo
  {author} {\bibfnamefont {G.~A.}\ \bibnamefont {Steele}},\ }\href {\doibase
  10.1038/s41534-018-0088-9} {\bibfield  {journal} {\bibinfo  {journal} {npj
  Quantum Information}\ }\textbf {\bibinfo {volume} {4}},\ \bibinfo {pages}
  {38} (\bibinfo {year} {2018})}\BibitemShut {NoStop}%
\bibitem [{\citenamefont {Hu}\ \emph {et~al.}(2011)\citenamefont {Hu},
  \citenamefont {Ge}, \citenamefont {Chen}, \citenamefont {Yang},\ and\
  \citenamefont {Chen}}]{PhysRevA.84.012329}%
  \BibitemOpen
  \bibfield  {author} {\bibinfo {author} {\bibfnamefont {Y.}~\bibnamefont
  {Hu}}, \bibinfo {author} {\bibfnamefont {G.-Q.}\ \bibnamefont {Ge}}, \bibinfo
  {author} {\bibfnamefont {S.}~\bibnamefont {Chen}}, \bibinfo {author}
  {\bibfnamefont {X.-F.}\ \bibnamefont {Yang}}, \ and\ \bibinfo {author}
  {\bibfnamefont {Y.-L.}\ \bibnamefont {Chen}},\ }\href {\doibase
  10.1103/PhysRevA.84.012329} {\bibfield  {journal} {\bibinfo  {journal} {Phys.
  Rev. A}\ }\textbf {\bibinfo {volume} {84}},\ \bibinfo {pages} {012329}
  (\bibinfo {year} {2011})}\BibitemShut {NoStop}%
\bibitem [{\citenamefont {Wang}\ \emph {et~al.}(2024)\citenamefont {Wang},
  \citenamefont {Nori},\ and\ \citenamefont {Xiang}}]{PhysRevLett.132.053601}%
  \BibitemOpen
  \bibfield  {author} {\bibinfo {author} {\bibfnamefont {B.}~\bibnamefont
  {Wang}}, \bibinfo {author} {\bibfnamefont {F.}~\bibnamefont {Nori}}, \ and\
  \bibinfo {author} {\bibfnamefont {Z.-L.}\ \bibnamefont {Xiang}},\ }\href
  {\doibase 10.1103/PhysRevLett.132.053601} {\bibfield  {journal} {\bibinfo
  {journal} {Phys. Rev. Lett.}\ }\textbf {\bibinfo {volume} {132}},\ \bibinfo
  {pages} {053601} (\bibinfo {year} {2024})}\BibitemShut {NoStop}%
\bibitem [{\citenamefont {Xu}\ \emph {et~al.}(2026)\citenamefont {Xu},
  \citenamefont {Sun},\ and\ \citenamefont {He}}]{vph8-186g}%
  \BibitemOpen
  \bibfield  {author} {\bibinfo {author} {\bibfnamefont {Y.}~\bibnamefont
  {Xu}}, \bibinfo {author} {\bibfnamefont {F.-X.}\ \bibnamefont {Sun}}, \ and\
  \bibinfo {author} {\bibfnamefont {Q.}~\bibnamefont {He}},\ }\href {\doibase
  10.1103/vph8-186g} {\bibfield  {journal} {\bibinfo  {journal} {Phys. Rev. A}\
  }\textbf {\bibinfo {volume} {113}},\ \bibinfo {pages} {023717} (\bibinfo
  {year} {2026})}\BibitemShut {NoStop}%
\bibitem [{\citenamefont {Blumenstein}\ \emph {et~al.}(2011)\citenamefont
  {Blumenstein}, \citenamefont {Schäfer}, \citenamefont {Mietke},
  \citenamefont {Meyer}, \citenamefont {Dollinger}, \citenamefont {Lochner},
  \citenamefont {Cui}, \citenamefont {Patthey}, \citenamefont {Matzdorf},\ and\
  \citenamefont {Claessen}}]{RN95}%
  \BibitemOpen
  \bibfield  {author} {\bibinfo {author} {\bibfnamefont {C.}~\bibnamefont
  {Blumenstein}}, \bibinfo {author} {\bibfnamefont {J.}~\bibnamefont
  {Schäfer}}, \bibinfo {author} {\bibfnamefont {S.}~\bibnamefont {Mietke}},
  \bibinfo {author} {\bibfnamefont {S.}~\bibnamefont {Meyer}}, \bibinfo
  {author} {\bibfnamefont {A.}~\bibnamefont {Dollinger}}, \bibinfo {author}
  {\bibfnamefont {M.}~\bibnamefont {Lochner}}, \bibinfo {author} {\bibfnamefont
  {X.~Y.}\ \bibnamefont {Cui}}, \bibinfo {author} {\bibfnamefont
  {L.}~\bibnamefont {Patthey}}, \bibinfo {author} {\bibfnamefont
  {R.}~\bibnamefont {Matzdorf}}, \ and\ \bibinfo {author} {\bibfnamefont
  {R.}~\bibnamefont {Claessen}},\ }\href {\doibase 10.1038/nphys2051}
  {\bibfield  {journal} {\bibinfo  {journal} {Nature Physics}\ }\textbf
  {\bibinfo {volume} {7}},\ \bibinfo {pages} {776} (\bibinfo {year}
  {2011})}\BibitemShut {NoStop}%
\bibitem [{\citenamefont {Wang}\ \emph {et~al.}(2022)\citenamefont {Wang},
  \citenamefont {Yu}, \citenamefont {Kwan}, \citenamefont {Jia}, \citenamefont
  {Lei}, \citenamefont {Klemenz}, \citenamefont {Cevallos}, \citenamefont
  {Singha}, \citenamefont {Devakul}, \citenamefont {Watanabe}, \citenamefont
  {Taniguchi}, \citenamefont {Sondhi}, \citenamefont {Cava}, \citenamefont
  {Schoop}, \citenamefont {Parameswaran},\ and\ \citenamefont {Wu}}]{RN96}%
  \BibitemOpen
  \bibfield  {author} {\bibinfo {author} {\bibfnamefont {P.}~\bibnamefont
  {Wang}}, \bibinfo {author} {\bibfnamefont {G.}~\bibnamefont {Yu}}, \bibinfo
  {author} {\bibfnamefont {Y.~H.}\ \bibnamefont {Kwan}}, \bibinfo {author}
  {\bibfnamefont {Y.}~\bibnamefont {Jia}}, \bibinfo {author} {\bibfnamefont
  {S.}~\bibnamefont {Lei}}, \bibinfo {author} {\bibfnamefont {S.}~\bibnamefont
  {Klemenz}}, \bibinfo {author} {\bibfnamefont {F.~A.}\ \bibnamefont
  {Cevallos}}, \bibinfo {author} {\bibfnamefont {R.}~\bibnamefont {Singha}},
  \bibinfo {author} {\bibfnamefont {T.}~\bibnamefont {Devakul}}, \bibinfo
  {author} {\bibfnamefont {K.}~\bibnamefont {Watanabe}}, \bibinfo {author}
  {\bibfnamefont {T.}~\bibnamefont {Taniguchi}}, \bibinfo {author}
  {\bibfnamefont {S.~L.}\ \bibnamefont {Sondhi}}, \bibinfo {author}
  {\bibfnamefont {R.~J.}\ \bibnamefont {Cava}}, \bibinfo {author}
  {\bibfnamefont {L.~M.}\ \bibnamefont {Schoop}}, \bibinfo {author}
  {\bibfnamefont {S.~A.}\ \bibnamefont {Parameswaran}}, \ and\ \bibinfo
  {author} {\bibfnamefont {S.}~\bibnamefont {Wu}},\ }\href {\doibase
  10.1038/s41586-022-04514-6} {\bibfield  {journal} {\bibinfo  {journal}
  {Nature}\ }\textbf {\bibinfo {volume} {605}},\ \bibinfo {pages} {57}
  (\bibinfo {year} {2022})}\BibitemShut {NoStop}%
\bibitem [{\citenamefont {Cui}\ \emph {et~al.}(2026)\citenamefont {Cui},
  \citenamefont {Yeh}, \citenamefont {Hulet}, \citenamefont {Pu}, \citenamefont
  {Giamarchi},\ and\ \citenamefont
  {Guan}}]{cui2026tomonagaluttingerliquidtheoryonedimensional}%
  \BibitemOpen
  \bibfield  {author} {\bibinfo {author} {\bibfnamefont {H.-Y.}\ \bibnamefont
  {Cui}}, \bibinfo {author} {\bibfnamefont {Y.-H.}\ \bibnamefont {Yeh}},
  \bibinfo {author} {\bibfnamefont {R.~G.}\ \bibnamefont {Hulet}}, \bibinfo
  {author} {\bibfnamefont {H.}~\bibnamefont {Pu}}, \bibinfo {author}
  {\bibfnamefont {T.}~\bibnamefont {Giamarchi}}, \ and\ \bibinfo {author}
  {\bibfnamefont {X.-W.}\ \bibnamefont {Guan}},\ }\href
  {https://arxiv.org/abs/2603.13958} {\enquote {\bibinfo {title}
  {Tomonaga-luttinger liquid theory for one-dimensional attractive fermi
  gases},}\ } (\bibinfo {year} {2026}),\ \Eprint
  {http://arxiv.org/abs/2603.13958} {arXiv:2603.13958 [cond-mat.quant-gas]}
  \BibitemShut {NoStop}%
\bibitem [{\citenamefont {M\"unstermann}\ \emph {et~al.}(2000)\citenamefont
  {M\"unstermann}, \citenamefont {Fischer}, \citenamefont {Maunz},
  \citenamefont {Pinkse},\ and\ \citenamefont {Rempe}}]{PhysRevLett.84.4068}%
  \BibitemOpen
  \bibfield  {author} {\bibinfo {author} {\bibfnamefont {P.}~\bibnamefont
  {M\"unstermann}}, \bibinfo {author} {\bibfnamefont {T.}~\bibnamefont
  {Fischer}}, \bibinfo {author} {\bibfnamefont {P.}~\bibnamefont {Maunz}},
  \bibinfo {author} {\bibfnamefont {P.~W.~H.}\ \bibnamefont {Pinkse}}, \ and\
  \bibinfo {author} {\bibfnamefont {G.}~\bibnamefont {Rempe}},\ }\href
  {\doibase 10.1103/PhysRevLett.84.4068} {\bibfield  {journal} {\bibinfo
  {journal} {Phys. Rev. Lett.}\ }\textbf {\bibinfo {volume} {84}},\ \bibinfo
  {pages} {4068} (\bibinfo {year} {2000})}\BibitemShut {NoStop}%
\bibitem [{\citenamefont {Yang}\ \emph {et~al.}(2021)\citenamefont {Yang},
  \citenamefont {Yin}, \citenamefont {Wen}, \citenamefont {Ji},\ and\
  \citenamefont {Sun}}]{PhysRevA.104.053313}%
  \BibitemOpen
  \bibfield  {author} {\bibinfo {author} {\bibfnamefont {M.-Y.}\ \bibnamefont
  {Yang}}, \bibinfo {author} {\bibfnamefont {H.-H.}\ \bibnamefont {Yin}},
  \bibinfo {author} {\bibfnamefont {L.}~\bibnamefont {Wen}}, \bibinfo {author}
  {\bibfnamefont {A.-C.}\ \bibnamefont {Ji}}, \ and\ \bibinfo {author}
  {\bibfnamefont {Q.}~\bibnamefont {Sun}},\ }\href {\doibase
  10.1103/PhysRevA.104.053313} {\bibfield  {journal} {\bibinfo  {journal}
  {Phys. Rev. A}\ }\textbf {\bibinfo {volume} {104}},\ \bibinfo {pages}
  {053313} (\bibinfo {year} {2021})}\BibitemShut {NoStop}%
\bibitem [{\citenamefont {Gao}\ \emph {et~al.}(2020)\citenamefont {Gao},
  \citenamefont {Schlawin}, \citenamefont {Buzzi}, \citenamefont {Cavalleri},\
  and\ \citenamefont {Jaksch}}]{PhysRevLett.125.053602}%
  \BibitemOpen
  \bibfield  {author} {\bibinfo {author} {\bibfnamefont {H.}~\bibnamefont
  {Gao}}, \bibinfo {author} {\bibfnamefont {F.}~\bibnamefont {Schlawin}},
  \bibinfo {author} {\bibfnamefont {M.}~\bibnamefont {Buzzi}}, \bibinfo
  {author} {\bibfnamefont {A.}~\bibnamefont {Cavalleri}}, \ and\ \bibinfo
  {author} {\bibfnamefont {D.}~\bibnamefont {Jaksch}},\ }\href {\doibase
  10.1103/PhysRevLett.125.053602} {\bibfield  {journal} {\bibinfo  {journal}
  {Phys. Rev. Lett.}\ }\textbf {\bibinfo {volume} {125}},\ \bibinfo {pages}
  {053602} (\bibinfo {year} {2020})}\BibitemShut {NoStop}%
\bibitem [{\citenamefont {Chakraborty}\ and\ \citenamefont
  {Piazza}(2021)}]{PhysRevLett.127.177002}%
  \BibitemOpen
  \bibfield  {author} {\bibinfo {author} {\bibfnamefont {A.}~\bibnamefont
  {Chakraborty}}\ and\ \bibinfo {author} {\bibfnamefont {F.}~\bibnamefont
  {Piazza}},\ }\href {\doibase 10.1103/PhysRevLett.127.177002} {\bibfield
  {journal} {\bibinfo  {journal} {Phys. Rev. Lett.}\ }\textbf {\bibinfo
  {volume} {127}},\ \bibinfo {pages} {177002} (\bibinfo {year}
  {2021})}\BibitemShut {NoStop}%
\bibitem [{\citenamefont {Rao}\ and\ \citenamefont
  {Piazza}(2023)}]{PhysRevLett.130.083603}%
  \BibitemOpen
  \bibfield  {author} {\bibinfo {author} {\bibfnamefont {P.}~\bibnamefont
  {Rao}}\ and\ \bibinfo {author} {\bibfnamefont {F.}~\bibnamefont {Piazza}},\
  }\href {\doibase 10.1103/PhysRevLett.130.083603} {\bibfield  {journal}
  {\bibinfo  {journal} {Phys. Rev. Lett.}\ }\textbf {\bibinfo {volume} {130}},\
  \bibinfo {pages} {083603} (\bibinfo {year} {2023})}\BibitemShut {NoStop}%
\bibitem [{\citenamefont {Torre}\ \emph {et~al.}(2013)\citenamefont {Torre},
  \citenamefont {Diehl}, \citenamefont {Lukin}, \citenamefont {Sachdev},\ and\
  \citenamefont {Strack}}]{PhysRevA.87.023831}%
  \BibitemOpen
  \bibfield  {author} {\bibinfo {author} {\bibfnamefont {E.~G.~D.}\
  \bibnamefont {Torre}}, \bibinfo {author} {\bibfnamefont {S.}~\bibnamefont
  {Diehl}}, \bibinfo {author} {\bibfnamefont {M.~D.}\ \bibnamefont {Lukin}},
  \bibinfo {author} {\bibfnamefont {S.}~\bibnamefont {Sachdev}}, \ and\
  \bibinfo {author} {\bibfnamefont {P.}~\bibnamefont {Strack}},\ }\href
  {\doibase 10.1103/PhysRevA.87.023831} {\bibfield  {journal} {\bibinfo
  {journal} {Phys. Rev. A}\ }\textbf {\bibinfo {volume} {87}},\ \bibinfo
  {pages} {023831} (\bibinfo {year} {2013})}\BibitemShut {NoStop}%
\bibitem [{\citenamefont {Dmytruk}\ and\ \citenamefont
  {Schir\'o}(2021)}]{PhysRevB.103.075131}%
  \BibitemOpen
  \bibfield  {author} {\bibinfo {author} {\bibfnamefont {O.}~\bibnamefont
  {Dmytruk}}\ and\ \bibinfo {author} {\bibfnamefont {M.}~\bibnamefont
  {Schir\'o}},\ }\href {\doibase 10.1103/PhysRevB.103.075131} {\bibfield
  {journal} {\bibinfo  {journal} {Phys. Rev. B}\ }\textbf {\bibinfo {volume}
  {103}},\ \bibinfo {pages} {075131} (\bibinfo {year} {2021})}\BibitemShut
  {NoStop}%
\end{thebibliography}%

\onecolumngrid

\section{The two mode Rabi model}
In this section, we will give the detailed derivation for the two mode Rabi model proposed in the main text.

\subsection{Floquet engineering for the effective Hamiltonian}
Firstly, we derive a detailed Floquet design~\cite{PhysRevLett.132.113402,PRXQuantum.4.020329,PhysRevLett.124.190601} for the effective Hamiltonian of the model. We consider the original Hamiltonian expressed as follows
\begin{align}
	H=\tilde{\omega}_bb^\dagger b+\tilde{\omega}_cc^\dagger c+\dfrac{\Omega}{2}(\sigma_1^z+\sigma_2^z)+\lambda(b+b^\dagger)(\sigma_1^x+\sigma_2^x)+\eta(b^\dagger c+bc^\dagger)(\sigma_1^x+\sigma_2^x).
\end{align}

Here, the $\sigma_{x,y,z}$ represents for the Pauli matrix in the subspace spanned by the magnetic energy levels $\ket{m=m_0}$ and $\ket{m=m_0+1}$ with the same angular quantum number $l$, and the transition frequency is $\Omega$ between these two inner states. $\tilde{\omega}_{b(c)}\equiv\omega_{b(c)}-\omega_{p}$ is the detuning between the original frequency $\omega_{b(c)}$ of the cavity eigenmode $b(c)$ and the pumping frequency $\omega_{p}$, which satisfied $\tilde{\omega}_{b(c)}\ll\Omega$, i.e., the pumping laser is near resonant with the cavity mode $b(c)$ compared to the transition energy $\Omega$. And the coupling strength satisfies $\lambda\sim\sqrt{\tilde{\omega}_b\Omega}$. It also refers to the classical oscillator limit in the Rabi superradiant model ~\cite{PhysRevLett.115.180404,PhysRevA.101.063843}. The propagation of mode $c$ is perpendicular to mode $b$ but aligned to the pumping laser in z-direction, which leads to the change of angular momentum in the stimulated Lamman process. Totally, our model consists of a well known Rabi model and a special "full-quantum" Rabi model, where a classical pumping laser is replaced by a quantum optical mode $c$. As a result, the full-quantum Rabi frequency $\eta$ is rather weak, but we assume that the coupling strength $\eta\left<c\right>$ will be unneglectable if mode $c$ is macroscopic condensed in cavity, i.e., superradiance occurs in mode $c$.


We assume that the coupling between $b$ and $c$ can be easily modified by a phase modulation, thus we can apply the Floquet engineering to construct the effective Hamiltonian, giving as
\begin{align}
	U=\exp(-iH_{eff}t)=[\exp(-iH\tau)\exp(-i\bar{H}\tau)]^{t/(2\tau)}.
\end{align}
Here $\bar{H}\equiv \exp(i\pi c^\dagger c)H\exp(i\pi c^\dagger c)$ means a half wavelength shift of the cavity mode $c$. To introduce the effective Hamiltonian, we use the Baker-Campbell-Hausdorff (BCH) formula, writing as 
\begin{align}
	e^Ae^B=\exp(A+B+\dfrac{1}{2}[A,B]+\dfrac{1}{12}[A,[A,B]]-\dfrac{1}{12}[B,[A,B]]+\dots).
\end{align}
Then, we give the calculations by considering the replacement $A\to-iH\tau$ and $B\to-i\bar{H}\tau$ as follows
\begin{align}
	[-i\tau\eta(b^\dagger c+bc^\dagger)(\sigma_1^x+\sigma_2^x),-i\tau\tilde{\omega}_bb^\dagger b]&=-i2\tau\dfrac{\eta\tilde{\omega}_b\tau}{2}(ib^\dagger c-ibc^\dagger)(\sigma_1^x+\sigma_2^x)\\
	[-i\tau\eta(b^\dagger c+bc^\dagger)(\sigma_1^x+\sigma_2^x),-i\tau\tilde{\omega}_cc^\dagger c]&=-i2\tau\dfrac{\eta\tilde{\omega}_c\tau}{2}(ic^\dagger b-icb^\dagger)(\sigma_1^x+\sigma_2^x)\\
	[-i\tau\eta(b^\dagger c+bc^\dagger)(\sigma_1^x+\sigma_2^x),-i\tau\dfrac{\Omega}{2}(\sigma_1^z+\sigma_2^z)]&=-i2\tau\dfrac{\eta\Omega\tau}{2}(b^\dagger c+bc^\dagger)(-\sigma_1^y-\sigma_2^y)\\
	[-i\tau\eta(b^\dagger c+bc^\dagger)(\sigma_1^x+\sigma_2^x),-i\tau\lambda(b+b^\dagger)(\sigma_1^x+\sigma_2^x)]&=-i2\tau\dfrac{\eta\lambda\tau}{2}(ic-ic^\dagger)(\sigma_1^x+\sigma_2^x)^2.
\end{align}
It's found that the first three lines give the coupling between the modes $b$, $c$ and the spins, and the third line gives the largest contribution. In classic oscillator limit, the covariance term can be safely ignored as $b^\dagger c\sigma^{x,y}_{1,2}\approx\left<b\right>^*\left<c\right>\sigma^{x,y}_{1,2}$. If we choose $\dfrac{\eta\Omega\tau}{2}\ll\left|\tilde{\omega}_b-\tilde{\omega}_c\right|$, then the coupling between $b$ and $c$ are suppressed. Also, we assume that $\dfrac{\eta\Omega\tau}{2}\left<c\right>\ll\lambda$, then the coupling between mode $b$ and two spins can be safely ignored. In addition, we consider the region around the critical boundary of the superradiant region of mode $b$, giving $\left<b\right>\approx0$, so the coupling between mode $c$ and two spins are viewed as zero. Therefore, only the fourth term is retained, and we request that $\dfrac{\eta\lambda\tau}{2}\gg\tilde{\omega}_c$. That is to say, we hierarchy relation for these parameters as
\begin{align}\label{hierarchy}
	\tilde{\omega}_c\ll\dfrac{\eta\lambda\tau}{2}\ll\dfrac{\eta\Omega\tau}{2}\ll\left|\tilde{\omega}_b-\tilde{\omega}_c\right|.
\end{align}
Furthermore, the condition $\Omega\tau\lesssim1$ is demmanded to be satisfied, making the higher order commutators sufficiently small compared with the 1st order one. Eventually, the effectively Hamiltonian after the Floquet engineering can writes as
\begin{align}\label{H_eff}
	H_{eff}&=\tilde{\omega}_bb^\dagger b+\tilde{\omega}_cc^\dagger c+\dfrac{\Omega}{2}(\sigma_1^z+\sigma_2^z)+\lambda(b+b^\dagger)(\sigma_1^x+\sigma_2^x)+\dfrac{\eta\lambda\tau}{2}(ic-ic^\dagger)(\sigma_1^x+\sigma_2^x)^2\notag\\
	&\rightarrow\tilde{\omega}_bb^\dagger b+\tilde{\omega}_cc^\dagger c+\dfrac{\Omega}{2}(\sigma_1^z+\sigma_2^z)+\lambda_b(b+b^\dagger)(\sigma_1^x+\sigma_2^x)-\dfrac{\lambda_c}{2}(c+c^\dagger)(\sigma_1^x+\sigma_2^x)^2.
\end{align}

In the second line, we redefine $H_{eff}\rightarrow \exp(-i\dfrac{\pi}{2}c^\dagger c)H_{eff}\exp(i\dfrac{\pi}{2}c^\dagger c)$, $\lambda_b\equiv\lambda$ and $\lambda_c\equiv\eta\lambda\tau$. The Eq.~(\ref{H_eff}) is exactly the one in main text.

Moreover, as a more scalable and experimental-friendly system, the circuit QED platform also provides new possiblity to implement the above Hamiltonian in microwave frequency band. In particular, the three-mode spontaneous parametric down-conversion has been realized in circuit QED systems~\cite{PhysRevX.10.011011,PhysRevApplied.10.044019,PhysRevLett.114.090503}, which facilitates the architecture of Eq.~(\ref{H_eff}).

\subsection{Phase transition of the ground state}

According to the hierarchy relation~(\ref{hierarchy}), our two-mode Rabi Hamiltonian satisfies classical oscillator limit, where the mean-field approximation is sufficient to evaluate the basic property of the system. Thus, we replace the operator $b(c)$ into its expectation value as $b(c)\rightarrow\left<b(c)\right>$ and define $\left<b\right>\equiv B,\left<c\right>\equiv C$. It gives the ground state energy of the system as 
\begin{align}\label{E(B,C)}
	E(B,C)&=\tilde{\omega}_bB^2+\tilde{\omega}_cC^2-2\lambda_cC+\left<\dfrac{\Omega}{2}(\sigma_1^z+\sigma_2^z)+2\lambda_bB(\sigma_1^x+\sigma_2^x)-2\lambda_cC\sigma_1^x\sigma_2^x\right>\notag\\
	&=\tilde{\omega}_bB^2+\tilde{\omega}_cC^2-2\lambda_cC+\left<H_{spin}(B,C)\right>.
\end{align}
Here, we define Hamiltonian of the spin part as $H_{spin}(B,C)\equiv\dfrac{\Omega}{2}(\sigma_1^z+\sigma_2^z)+2\lambda_bB(\sigma_1^x+\sigma_2^x)-2\lambda_cC\sigma_1^x\sigma_2^x$, and $\left<\dots\right>$ means the expectation value in the ground state of the spin part, i.e., the minimum eigenvalue of the terms in angle brackets. Then we can expand the spin Hamiltonian in the total angular momentum basis as $\ket{S,m_S}\in\left\{\ket{1,1},\ket{1,0},\ket{1,-1},\ket{0,0}\right\}$.

\begin{align}
	H_{spin}(B,C)=\left(\begin{array}{cccc}
		\Omega&2\sqrt{2}\lambda_bB&-2\lambda_cC&0\\
		2\sqrt{2}\lambda_bB&-2\lambda_cC&2\sqrt{2}\lambda_bB&0\\
		-2\lambda_cC&2\sqrt{2}\lambda_bB&-\Omega&0\\
		0&0&0&2\lambda_cC
	\end{array}\right)
\end{align}

The structure of the energy in Eq.~(\ref{E(B,C)}) refers to the universal expression of free energy by mapping the order parameters $(B,C)$ into $(O_2,O_1)$. If we consider the case where the $b$ mode near critical the boundary between the normal phase and superradiant phase, where the expectation value $B\to0$, then we can apply the pertubative theory by writing $H_{spin}=H_0(C)+H'(B)$, where the unpertubative Hamiltonian $H_0(C)$ and the pertubative Hamiltonian $H'(B)$ can be expressed as
\begin{align}
	H_0(C)=\left(\begin{array}{cccc}
		\Omega&0&-2\lambda_cC&0\\
		0&-2\lambda_cC&0&0\\
		-2\lambda_cC&0&-\Omega&0\\
		0&0&0&2\lambda_cC
	\end{array}\right)\\
	H'(B)=\left(\begin{array}{cccc}
		0&2\sqrt{2}\lambda_bB&0&0\\
		2\sqrt{2}\lambda_bB&0&2\sqrt{2}\lambda_bB&0\\
		0&2\sqrt{2}\lambda_bB&0&0\\
		0&0&0&0
	\end{array}\right).
\end{align}
Firstly, we consider the unpertubative part $H_0$. The lowest and highest energy basis will be the superposition between $\ket{S=1,m_S=1}$ and $\ket{S=1,m_S=-1}$. Define the pairing order $\Delta_c\equiv2\lambda_cC$, we can easily check that $\Delta_c>0$ in order for approaching minimum value of Eq.~(\ref{E(B,C)})with the linear terms $-2\lambda_cC$. The highest and lowest eigenenergies are expressed as $E_\pm=\pm\sqrt{\Omega^2+\Delta_c^2}$ with the eigenbasis $\ket{+}=\cos\theta\ket{1,1}+\sin\theta\ket{1,-1}$ and $\ket{-}=-\sin\theta\ket{1,1}+\cos\theta\ket{1,-1}$, respectively, where the angle $\theta$ is defined as $\tan(2\theta)=-\dfrac{\Delta_c}{\Omega}$. Diagonalize $H_0$ with the basis $\left\{\ket{\pm},\ket{1,0},\ket{ 0,0}\right\}$, then the Hamiltonian in the new basis can be expressed as 
\begin{align}
	\tilde{H}_0(C)&=\left(\begin{array}{cccc}
		E_+&0&0&0\\
		0&-\Delta_c&0&0\\
		0&0&E_-&0\\
		0&0&0&\Delta_c
	\end{array}\right)\\
	\tilde{H}'(B)&=2\sqrt{2}\lambda_bB\left(\begin{array}{cccc}
		0&\sin\theta+\cos\theta&0&0\\
		\sin\theta+\cos\theta&0&\sin\theta-\cos\theta&0\\
		0&\sin\theta-\cos\theta&0&0\\
		0&0&0&0
	\end{array}\right).
\end{align}
Obviously, on the critical boundary where the expectation value $B$ increases from zero, the pertubative part will add the correction to the lowest eigenenergy $E_-$ with $E_-^{(2)}=8\lambda_b^2B^2(\sin\theta-\cos\theta)^2\dfrac{1}{E_-+\Delta_c}$. Therefore the lowest eigenvalue of the spin Hamiltonian $H_{spin}(B,C)$ is obtained as $E_-^s=E_-+E_-^{(2)}+\dots$. Thus, we can obtained the susceptibility of the mode $b$ according to the Eq.~(\ref{E(B,C)})
\begin{align}
	\dfrac{\partial E}{\partial(B^2)}\Bigg|_{B=0}&=\tilde{\omega}_b+\dfrac{8\lambda_b^2(\sin\theta-\cos\theta)^2}{E_-+\Delta_c}=\tilde{\omega}_b+8\lambda_b^2\dfrac{1-\sin(2\theta)}{E_-+\Delta_c}\notag\\
	&=\tilde{\omega}_b-8\lambda_b^2\dfrac{1+\Delta_c/\sqrt{\Omega^2+\Delta_c^2}}{\sqrt{\Omega^2+\Delta_c^2}-\Delta_c}\notag\\
	&=\tilde{\omega}_b-\dfrac{8\lambda_b^2}{\Omega^2}(\sqrt{\Omega^2+\Delta_c^2}+\dfrac{\Delta_c^2}{\sqrt{\Omega^2+\Delta_c^2}}+2\Delta_c).
\end{align}
And the critical point for the continuous phase transition can be obtained according to $\dfrac{\partial E}{\partial(B^2)}\Bigg|_{B=0}=0$. On the other hand, we can evaluate the second order differentiation as follows
\begin{align}
	\dfrac{\partial^2E}{\partial\Delta_c\partial(B^2)}\Bigg|_{B=0,\Delta_c=\Delta_{c0}}&=\dfrac{\partial}{\partial\Delta_c}\left(\dfrac{\partial E}{\partial(B^2)}\Bigg|_{B=0}\right)\Bigg|_{\Delta_c=\Delta_{c0}}\\
	&=-\dfrac{8\lambda_b^2}{\Omega^2}\left[2-\left(\dfrac{\Delta_{c0}}{\sqrt{\Omega^2+\Delta_{c0}^2}}\right)^3+\dfrac{3\Delta_{c0}}{\sqrt{\Omega^2+\Delta_{c0}^2}}\right].
\end{align}
Eventually, we can easily prove that $\dfrac{\partial^2E}{\partial\Delta_c\partial(B^2)}\Bigg|_{B=0,\Delta_c=\Delta_{c0}}<0$ for arbitrary initial pairing order $\Delta_{c0}$. By means of the criterion discussed in the main text, once the superradiance just occurs in mode $b$ or $\dfrac{\partial E}{\partial(B^2)}\Bigg|_{B=0}=0^-$, the expectation $B^2$ will be slightly larger from zero. Meanwhile, the pairing order $\Delta_c$ will also become larger from the initial $\Delta_{c0}>0$. That is to say, the order parameters will change as $[B^2=0,\Delta_c=\Delta_{c0}]\rightarrow[B^2=\delta(B^2),\Delta_c=\Delta_{c0}+\delta\Delta_c]$ with $\delta(B^2),\delta\Delta_c>0$, if the susceptibility $\dfrac{\partial E}{\partial(B^2)}\Bigg|_{B=0}$ changes from $0^+$ to $0^-$.

In order for further verification of our criterion, we consider a rather weak cross Kerr term $H_{\mathcal{K}}=\chi_3B^2C^2+\chi(B^4+C^4)$ in our system, which won't affect the Floquet engineering above. It’s also feasible and even tunable in circuit QED systems~\cite{RN92,PhysRevA.84.012329}. Taking this effect into account, we can modify the 2nd order differentiation as 
\begin{align}
	\dfrac{\partial E}{\partial(B^2)}\Bigg|_{B=0}&=\tilde{\omega}_b-\dfrac{8\lambda_b^2}{\Omega^2}(\sqrt{\Omega^2+\Delta_c^2}+\dfrac{\Delta_c^2}{\sqrt{\Omega^2+\Delta_c^2}}+2\Delta_c)+\dfrac{\chi_3\Delta_{c}^2}{4\lambda_c^2}\\
	\dfrac{\partial^2E}{\partial\Delta_c\partial(B^2)}\Bigg|_{B=0,\Delta_c=\Delta_{c0}}&=\dfrac{\chi_3\Delta_{c0}}{2\lambda_c^2}-\dfrac{8\lambda_b^2}{\Omega^2}\left[2-\left(\dfrac{\Delta_{c0}}{\sqrt{\Omega^2+\Delta_{c0}^2}}\right)^3+\dfrac{3\Delta_{c0}}{\sqrt{\Omega^2+\Delta_{c0}^2}}\right].
\end{align}

In another aspect, the 2nd order differentiation is expressed as
\begin{align}
	\dfrac{\partial^2E}{\partial\Delta_c^2}\Bigg|_{B=0,\Delta_c=\Delta_{c0}}=\dfrac{\tilde{\omega}_c}{2\lambda_c^2}-\dfrac{1}{\sqrt{\Omega^2+\Delta_{c0}^2}}+\dfrac{\Delta_{c0}^2}{(\Omega^2+\Delta_{c0}^2)^{3/2}}+\dfrac{3\chi\Delta_{c0}^2}{4\lambda_c^4}.
\end{align}


We give a linear fitting as $\Delta_c=\Delta_{c0}-\dfrac{\partial^2E/\partial\Delta_c\partial(B^2)}{\partial^2E/\partial\Delta_c^2}\Bigg|_{B=0,\Delta_c=\Delta_{c0}}\times B^2$, matching well with the numerical results in Fig. 1 of main text.

\section{The Fermi Dicke model}
In this section, we will give the detailed derivation for the Fermi Dicke model mentioned in the main text.

\subsection{Derive the effective Hamiltonian}
The original Hamiltonian can be divided into serveral parts, given as $H=H_0+H_{FA}+H_{FF}$. $H_0$ is the energy of the eigenmode of the cavity and the free Fermi gas. $H_{FA}$ and $H_{FF}$ are the atom-cavity interaction and the atom-atom scattering terms respectively.
\begin{align}
	H_0&=\omega_ca^\dagger a+\sum_{\sigma=\downarrow,\uparrow}\omega_{\sigma,e}\ket{\sigma,e}\bra{\sigma,e}+h(\ket{\uparrow,g}\bra{\uparrow,g}-\ket{\downarrow,g}\bra{\downarrow,g}).\\
	H_{FA}&=-\Omega(\sigma_{1\uparrow}^++\sigma_{2\downarrow}^+)e^{-i\omega_pt}cos(k_pz)-g(\sigma_{1\downarrow}^+a+\sigma_{2\uparrow}^+a)cos(k_cx)+H.c.
\end{align}

Here, the $\sigma_{1\sigma}^+=\ket{\downarrow,e}\bra{\sigma,g}$ and, $\sigma_{2\sigma}^+=\ket{\uparrow,e}\bra{\sigma,g}$. By means of the unitary transformation $U(t)=\exp[i(\sum_{\sigma=\uparrow,\downarrow}\ket{\sigma,e}\bra{\sigma,e}+a^\dagger a)\omega_{p}t]$, the single particle Hamiltonian $H_{single}\equiv H_0+H_{FA}$ in the rotation frame can be written as 
\begin{align}
	\tilde{H}_{single}&=\tilde{\omega}a^\dagger a+\delta_{\uparrow}\ket{\uparrow,e}\bra{\uparrow,e}+\delta_{\downarrow}\ket{\downarrow,e}\bra{\downarrow,e}+h(\ket{\uparrow,g}\bra{\uparrow,g}-\ket{\downarrow,g}\bra{\downarrow,g})\notag\\
	&-[\Omega(\sigma_{1\uparrow}^++\sigma_{2\downarrow}^+)cos(k_pz)+g(\sigma_{1\downarrow}^+a+\sigma_{2\uparrow}^+a)cos(k_cx)+H.c.],
\end{align}
where $\tilde{\omega}\equiv\omega-\omega_{p}$ and $\delta_{\sigma}\equiv\omega_{\sigma,e}-\omega_{p}$ stand for cavity and atom detunings against the pump laser, respectively. For simplicity, we only the Zeeman splitting energy is much smaller than the detunings, satisfying $h, \left|\omega_{\uparrow,e}-\omega_{\downarrow,e}\right|\ll\tilde{\omega},\delta_{\sigma}$. Without loss of generality, we apply the approximation $\delta_{\uparrow}\approx\delta_{\downarrow}=\delta$ and $h\approx0$ in the following calculation. 

Considering the dynamic evolution of the eigenbasis of the fermion inner states $\left\{\ket{\sigma,g},\ket{\sigma,e}\right\}$, we can write the equations according to the Schr\"{o}dinger equations driven by the Hamiltonian $\tilde{H}_I$,
\begin{align}
	&i\dfrac{d}{dt}\ket{\downarrow,e}=\delta\ket{\downarrow,e}-\Omega cos(k_pz)\ket{\uparrow,g}-g\cos(k_cx)a^\dagger\ket{\downarrow,g}.\\
	&i\dfrac{d}{dt}\ket{\uparrow,e}=\delta\ket{\uparrow,e}-\Omega cos(k_pz)\ket{\downarrow,g}-g\cos(k_cx)a^\dagger\ket{\uparrow,g}.\\
	&i\dfrac{d}{dt}\ket{\downarrow,g}=-\Omega cos(k_pz)\ket{\uparrow,e}-g\cos(k_cx)a\ket{\downarrow,e}.\\
	&i\dfrac{d}{dt}\ket{\uparrow,g}=-\Omega cos(k_pz)\ket{\downarrow,e}-g\cos(k_cx)a\ket{\uparrow,e}.
\end{align}

Assuming that the detuning $\delta$ is large enough, i.e., $\delta\gg g$, the system maintains in the subspace $\{\ket{\downarrow,g},\ket{\uparrow,g}\}$ with the probability approaching 1. Therefore we can apply adiabatic elimination for the excited states $\ket{e}$, and the effective dynamic evolution for the ground state can be obtained approximately as
\begin{align}
	i\dfrac{d}{dt}\ket{\downarrow,g}=&-\dfrac{1}{\delta}[\Omega^2\cos^2(k_pz)\ket{\downarrow,g}+g^2\cos^2(k_cx)a^\dagger a\ket{\downarrow,g}+\Omega g(a+a^\dagger)\cos(k_pz)\cos(k_cx)\ket{\uparrow,g}]\notag\\
	i\dfrac{d}{dt}\ket{\uparrow,g}=&-\dfrac{1}{\delta}[\Omega^2\cos^2(k_pz)\ket{\uparrow,g}+g^2\cos^2(k_cx)a^\dagger a\ket{\uparrow,g}+\Omega g(a+a^\dagger)\cos(k_pz)\cos(k_cx)\ket{\downarrow,g}].
\end{align}

We emphasize that we didn't raise a claim that $\delta\gg\Omega$ for adiabatic elimilation, because if we assume the pumping strength $\Omega$ is large enough, a periodic potential $V\sim\cos^2(k_pz)$ in z direction will be built, which suppresses the cold atoms in the wave valleys of the optical lattice, giving local small coupling strength satisfying $\delta\gg\Omega\cos(k_pz)$, making the adiabatic elimilation valid. Then, the effective Hamiltonian after the adiabatic elimination is expressed as follows:
\begin{align}
	\tilde{H}_{single}\approx\tilde{H}_{single}^{ad}=&[\tilde{\omega}-\dfrac{g^2}{\delta}\cos^2(k_cx)]a^\dagger a-\dfrac{g\Omega}{\delta}\cos(k_cx)\cos(k_pz)(a+a^\dagger)\sigma_x-\dfrac{\Omega^2}{\delta}\cos^2(k_pz).
\end{align}
Taking the kinetic energy of fermions into account, the total Hamiltonian can be written as
\begin{align}\label{H_tot}
	\tilde{H}_{single}&\approx\tilde{\omega}a^\dagger a+\sum_{\sigma,k}\dfrac{k^2}{2m}c_{\sigma,k}^\dagger c_{\sigma,k}-\dfrac{g^2}{4\delta}a^\dagger a\sum_{\sigma,k,s=\pm1}c_{\sigma,k}^\dagger c_{\sigma,k+2sk_c}\notag\\
	&-[\dfrac{g\Omega}{4\delta}(a+a^\dagger)\sum_{k,s,s'=\pm1}c_{\uparrow,k}^\dagger c_{\downarrow,k+sk_c+s'k_p}+H.c.]-\dfrac{\Omega^2}{4\delta}\sum_{\sigma,k,s=\pm1}c_{\sigma,k}^\dagger c_{\sigma,k+2sk_p}.
\end{align}
Here, the sign of "$\approx$" originates from the assumption that $\tilde{\omega}\approx\tilde{\omega}-\dfrac{g^2}{2\delta}$ by consider a rather small coupling strength $g$ reasonably. In this work, we concentrate on the case where the pump laser is strong enough, so that the Fermi gas is highly localized in $z$ direction. Since the wave vector of the cavity mode is perpendicular to the pump laser, the fermi gas can be viewed as a 2D ensemble in $x$-$y$ plane, and the dispersive in pump direction $z$ is flat. This can be given with an equivalent description as $\dfrac{\Omega^2}{\delta E_r}\gg1$, where $E_r=\dfrac{k_c^2}{2m}$ is the recoil energy, and the last term is negligible. As a technique consideration, we change the cavity-atom coupling strength $g$ to $g/\sqrt{2N}$, where $\sqrt{2N}$ in the denominator is also called a renormalization factor, making the results independent of the macroscopic atom number $2N$. And we also assume that $\dfrac{g^2}{4\delta}\ll\tilde{\omega},E_r$ so that the third term on the right hand side of Eq.~(\ref{H_tot}) can hardly affect the background potential of the Fermi gas. In aspect of perturbation theory, we consider the strength coefficient in every order of $a(a^\dagger)$, the contribution from the fourth term will also dominate the one from the third term. Based on the considerations above, the Hamiltonian can be further simplified as 
\begin{align}
	\tilde{H}_{single}&=\tilde{\omega}a^\dagger a+\sum_{\sigma,k}\dfrac{k^2}{2m}c_{\sigma,k}^\dagger c_{\sigma,k}-[\dfrac{g\Omega}{4\delta\sqrt{2N}}(a+a^\dagger)\sum_{k,s=\pm1}c_{\uparrow,k}^\dagger c_{\downarrow,k+sk_c}+H.c.].
\end{align}

Then, the fermionic interaction can be expressed as 
\begin{align}
	H_{FF}&=\dfrac{U}{2N}\sum_{k_1,k_2,q}c^\dagger_{\uparrow,k_1+q}c^\dagger_{\downarrow,k_2-q}c_{\downarrow,k_2}c_{\uparrow,k_1}\notag\\
	&\approx\dfrac{U}{2N}\sum_{k_1,k_2}c^\dagger_{\uparrow,k_1}c^\dagger_{\downarrow,-k_1}c_{\downarrow,-k_2}c_{\uparrow,k_2}\notag\\
	&\approx -\Delta\sum_{k}c^\dagger_{\uparrow,k}c^\dagger_{\downarrow,-k}-\Delta^*\sum_{k}c_{\downarrow,-k}c_{\uparrow,k}-\dfrac{2N\Delta^2}{U}.
\end{align}

In the second line, we only consider the backward scattering process where the total momentum is equal to zero. By means of Mean-field Approximation, we give the s-wave scattering interaction formula with the order parameter defined as $\Delta\equiv-\dfrac{U}{2N}\left<\sum_{k}c_{\downarrow,-k}c_{\uparrow,k}\right>$, which describes the conventional BCS superconductor band gap.

\subsection{The superradiant phase and superfluid phase}

According to the Landau's theory, the order parameter is obtained by the minimum of the free energy $F=U-TS$. In this paper, we focus on the zero temperature limit, thus only the internal energy is what we concern. In the thermodynamic limit where the number of fermionic atom $2N\rightarrow\infty$, the expectation value of the optical mode satisfies $\left<a\right>\sim\sqrt{2N}$, and it will be a rather good approximation to replace the optical mode $a$ into its expectation value $\alpha$. Based on above consideration, the Free energy operator can write as

\begin{align}
	\hat{F}&=\tilde{\omega}\left|\alpha\right|^2+\sum_{\sigma,k}(\dfrac{k^2}{2m}-\mu)c_{\sigma,k}^\dagger c_{\sigma,k}-[\dfrac{g\Omega}{4\delta\sqrt{2N}}(\alpha+\alpha^*)\sum_{k,s=\pm1}c_{\uparrow,k}^\dagger c_{\downarrow,k+sk_c}+H.c.]\notag\\
	&+\Delta\sum_{k}c^\dagger_{\uparrow,k}c^\dagger_{\downarrow,-k}+\Delta^*\sum_{k}c_{\downarrow,-k}c_{\uparrow,k}-\dfrac{2N\Delta^2}{U}+2\mu N.
\end{align}
Here, we introduce the chemical potential $\mu(\alpha,\Delta)$ to keep the total atom number $2N$ conservative, i.e., $\dfrac{\partial F}{\partial\mu}=0$. The global minimum of the expectation value $F(\alpha,\Delta)=\left<\hat{F}(\alpha,\Delta)\right>$ in the parameter plane $(\alpha,\Delta)$ satisfies $\dfrac{\partial F}{\partial\alpha}=\dfrac{\partial F}{\partial\Delta}=0$.
\begin{align}
	\dfrac{\partial F}{\partial\alpha}=\dfrac{\partial F}{\partial\alpha}\Bigg|_\mu+\dfrac{\partial F}{\partial\mu}\dfrac{\partial\mu}{\partial\alpha}=\dfrac{\partial F}{\partial\alpha}\Bigg|_\mu+\dfrac{\partial\mu}{\partial\alpha}(2N-\sum_{\sigma,k}\left<c_{\sigma,k}^\dagger c_{\sigma,k}\right>)=\dfrac{\partial F}{\partial\alpha}\Bigg|_\mu.
\end{align}
Here, we use the relation $2N=\sum_{\sigma,k}\left<c_{\sigma,k}^\dagger c_{\sigma,k}\right>$ in the last equation, or we can directly use $\dfrac{\partial F}{\partial\mu}=0$. And the similar relation can be obtained as $\dfrac{\partial F}{\partial\Delta}=\dfrac{\partial F}{\partial\Delta}\Bigg|_\mu$. According to the former research works on the BCS superfluid and superradiance~\cite{PhysRevA.104.053313}, we expect that the larger coupling strength $g\Omega/\delta$ will lead to the superradiant phase transition, and the attractive strength $\left|U\right|$ ($U<0$) decides the nonzero superfluid order parameter $\Delta$.

In order to understand the basic physics picture of this model, we first consider the following two cases: $\Omega\neq0, U=0$ and $\Omega=0, U\neq0$. The former case will give the the non-interaction fermionic superradiance after a critical pumping strength $\Omega_c^0$, which is well studied by ~\cite{PhysRevLett.112.143002,PhysRevLett.112.143003,PhysRevLett.112.143004}. The latter case leads to a conventional 1D BCS superconductor, and the order parameter $\Delta$ can be analytically decided by
\begin{align}
	1+\dfrac{U}{4N}\sum_k\dfrac{1}{\sqrt{(\epsilon_k-\mu)^2+\Delta^2}}=0.
\end{align}
In 2D and 3D systems, the short-range attractive interaction will give ultraviolet divergency of the summation, so the renormalization equation must be introduced to substitute the non-physical scattering strength $U$ into physical scattering length $a_s$. While for 1D system, the summation is converged as $k\rightarrow\infty$, and the renormalization equation isn't necessary, so we simply keep the scattering strength $U$ in this paper. Although that the superfluid phase will be strongly suppressed by quantum fluctuation in 1D and finite temperature case, where the Luttinger liquid theory is supposed to be applied to describe such system, we will focus on the 1D zero temperature case where the basic physics is still important and instructive to higher dimensional and finite temperature case. 

Therefore, we expect that both the superradiance and BCS order can be observed in our system. On one hand, the non-zero BCS order parameter will influence the critical coupling strength of superradiance. On the another hand, the superradiance will also influence superconductor energy gap $\Delta$.

\subsection{The critical phenomena of the superradiant phase transition}


In this section, we calculate the critical coupling strength for superradiance in presence of a nonzero $\Delta$.

Consider the critical case, where the superradiant order parameter $\left|\alpha\right|/\sqrt{2N}=0^+$, we can view the atom-light coupling term as the pertubative term, giving $H=H_0+H'$ 
\begin{align}
	H_0&=\tilde{\omega}\left|\alpha\right|^2+\sum_{\sigma,k}(\dfrac{k^2}{2m}-\mu)c_{\sigma,k}^\dagger c_{\sigma,k}-\Delta\sum_{k}c^\dagger_{\uparrow,k}c^\dagger_{\downarrow,-k}-\Delta^*\sum_{k}c_{\downarrow,-k}c_{\uparrow,k}-\dfrac{2N\Delta^2}{U}=\tilde{\omega}\left|\alpha\right|^2+H_0^f.\\
	H'&=-[\dfrac{g\Omega}{4\delta\sqrt{2N}}(\alpha+\alpha^*)\sum_{k,s=\pm1}c_{\uparrow,k}^\dagger c_{\downarrow,k+sk_c}+H.c.],
\end{align}
where the optical mode $a$ is replaced into the expectation value $\alpha$. In the expression of $H_0$, the  Numbu representation can be applied to express the fermionic part $H_0^f$ into a more compact form as 
\begin{align}
	H_0^f&\equiv\sum_{\sigma,k}(\dfrac{k^2}{2m}-\mu)c_{\sigma,k}^\dagger c_{\sigma,k}-\Delta\sum_{k}c^\dagger_{\uparrow,k}c^\dagger_{\downarrow,-k}-\Delta\sum_{k}c_{\downarrow,-k}c_{\uparrow,k}-\dfrac{2N\Delta^2}{U}\notag\\
	&=\sum_{k}\left(\begin{array}{cc}
		c_{\uparrow,k}^\dagger&
		c_{\downarrow,-k}
	\end{array}\right)\left(\begin{array}{cc}
		\dfrac{k^2}{2m}-\mu&-\Delta\\
		-\Delta&-(\dfrac{k^2}{2m}-\mu)
	\end{array}\right)\left(\begin{array}{c}
		c_{\uparrow,k}\\
		c_{\downarrow,-k}^\dagger
	\end{array}\right)+\sum_{k}(\dfrac{k^2}{2m}-\mu)-\dfrac{2N\Delta^2}{U}.
\end{align}
Here, we assume the order parameter $\Delta$ is real, because the original Hamiltonian maintains $U(1)$ symmetry before Mean-field approximation, indicating the total number conservation. The occurance of superfluid will breaks the continuous $U(1)$ symmetry and gives an fixed $\left|\Delta\right|$ but with arbitrary phase $\Delta/\left|\Delta\right|$. By means of Bogoliubov transformation, We can diagonalize the Hamiltonian into 
\begin{align}\label{interaction_fermi}
	H_0^f&=\sum_{k}\mathcal{E}_k(\alpha_k^\dagger\alpha_k+\beta_k^\dagger\beta_k)+\sum_{k}[(\dfrac{k^2}{2m}-\mu)-\mathcal{E}_k]-\dfrac{2N\Delta^2}{U}.
\end{align} 
Here the excited energy of the Bogoliubov quasi-particle is given by $\mathcal{E}_k=\sqrt{(\dfrac{k^2}{2m}-\mu)^2+\Delta^2}$, and $\alpha_k,\beta_k$ are the annihilation operator of the quasi-particle, writing as
\begin{align}
	c_{\uparrow,k}&=u_k\alpha_k+v_k\beta_k^\dagger\\
	c_{\downarrow,-k}^\dagger&=-v_k\alpha_k+u_k\beta_k^\dagger.
\end{align}
The coefficients are given as $u_k^2=\dfrac{1}{2}\left(1+\dfrac{k^2/2m-\mu}{\mathcal{E}_k}\right)$ and $v_k^2=\dfrac{1}{2}\left(1-\dfrac{k^2/2m-\mu}{\mathcal{E}_k}\right)$, where the sign of $u_k$ and $v_k$ is decided by the sign of $\Delta$. In the $\Delta\to0$ limit, on one hand, $\beta_k^\dagger=c_{\uparrow,k}$ and $\alpha_k^\dagger=-c_{\downarrow,-k}$ with the $k^2/2m-\mu<0$, which means the hole excitations under the fermi sea. On the other hand, $\alpha_k^\dagger=c_{\uparrow,k}^\dagger$, and $\beta_k^\dagger=c_{\downarrow,-k}^\dagger$ with with the $k^2/2m-\mu>0$, which means the particle excitations above the fermi sea. The ground state of the Hamiltonian~(\ref{interaction_fermi}) is expressed as the well-known BCS ground state
\begin{align}
	\ket{\Psi^{BCS}}=\prod_k\ket{\Psi^{BCS}_k}=\prod_k(u_k+v_kc_{\uparrow,k}^\dagger c_{\downarrow,-k}^\dagger)\ket{0}.
\end{align}
Similar to the real vaccum state $\ket{0}$, the BCS ground state $\ket{\Psi^{BCS}}$ refers to the vaccum of the Bogoliubov quasi-particles, satisfying $\alpha_k\ket{\Psi^{BCS}}=\beta_k\ket{\Psi^{BCS}}=0$. And $\ket{\Psi^{BCS}_k}\equiv(u_k+v_kc_{\uparrow,k}^\dagger c_{\downarrow,-k}^\dagger)\ket{0}$ represents the $k$ component of the BCS ground state. Additionally, we can also check that
\begin{align}
	\alpha_{k_0}^\dagger\ket{\Psi^{BCS}}&=c_{\uparrow,k_0}^\dagger\prod_{k\neq k_0}\ket{\Psi^{BCS}_k}\\
	\beta_{k_0}^\dagger\ket{\Psi^{BCS}}&=c_{\downarrow,-k_0}^\dagger\prod_{k\neq k_0}\ket{\Psi^{BCS}_k}\\
	\alpha_{k_0}^\dagger\beta_{k_0}^\dagger\ket{\Psi^{BCS}}&=(-\nu_{k_0}+u_{k_0}c_{\uparrow,k_0}^\dagger c_{\downarrow,-k_0}^\dagger)\prod_{k\neq k_0}\ket{\Psi^{BCS}_k},
\end{align}
giving the single excited state and double-excited state with excited energies $\mathcal{E}_k$ and $2\mathcal{E}_k$ respectively.

By means of the creation and annihilation operator of the Bogoliubov quasi-particles, we can expand the pertubative Hamiltonian $H'$. By consider the all the second order contribution $\sum_{m}\bra{\Psi^{BCS}}H'\ket{m}\bra{m}H'\ket{\Psi^{BCS}}$, where $\ket{m}$ stands for an arbitrary accessible intermediate state during the scattering process. We consider the term
\begin{align}
	&c_{\uparrow,k}^\dagger c_{\downarrow,k+sk_c}=\left(u_k\alpha_k+v_k\beta_k^\dagger\right)^\dagger\left(-v_{-(k+sk_c)}\alpha_{-(k+sk_c)}+u_{-(k+sk_c)}\beta_{-(k+sk_c)}^\dagger\right)^\dagger\notag\\
	=&-u_kv_{-(k+sk_c)}\alpha_k^\dagger\alpha_{-(k+sk_c)}^\dagger+u_ku_{-(k+sk_c)}\alpha_k^\dagger\beta_{-(k+sk_c)}-v_kv_{-(k+sk_c)}\beta_k\alpha_{-(k+sk_c)}^\dagger+v_ku_{-(k+sk_c)}\beta_k\beta_{-(k+sk_c)}\\
	&c_{\downarrow,-k}c_{\uparrow,-(k+sk_c)}^\dagger=\left(-v_k\alpha_k+u_k\beta_k^\dagger\right)^\dagger\left(u_{-(k+sk_c)}\alpha_{-(k+sk_c)}+v_{-(k+sk_c)}\beta_{-(k+sk_c)}^\dagger\right)^\dagger\notag\\
	=&-v_ku_{-(k+sk_c)}\alpha_k^\dagger\alpha_{-(k+sk_c)}^\dagger-v_kv_{-(k+sk_c)}\alpha_k^\dagger\beta_{-(k+sk_c)}+u_ku_{-(k+sk_c)}\beta_k\alpha_{-(k+sk_c)}^\dagger+u_kv_{-(k+sk_c)}\beta_k\beta_{-(k+sk_c)}
\end{align}
Obviously, only the first terms gives a nonzero contribution, expressed as
\begin{align}
	\left(c_{\uparrow,k}^\dagger c_{\downarrow,k+sk_c}+c_{\uparrow,-(k+sk_c)}^\dagger c_{\downarrow,-k}\right)\ket{\Psi^{BCS}}=\left(-u_kv_{-(k+sk_c)}+v_ku_{-(k+sk_c)}\right)c_{\uparrow,k}^\dagger c_{\uparrow,-(k+sk_c)}^\dagger\prod_{k'\neq k,-(k+sk_c)}\ket{\Psi^{BCS}_k}.
\end{align}
Here we have used the fermionic anti-commutation relation.

Similarly, the Hermitian terms $c_{\downarrow,k+sk_c}^\dagger c_{\uparrow,k}+c_{\downarrow,-k}^\dagger c_{\uparrow,-(k+sk_c)}$ will lead the counterpart contribution
\begin{align}
	\left(c_{\downarrow,k+sk_c}^\dagger c_{\uparrow,k}+c_{\downarrow,-k}^\dagger c_{\uparrow,-(k+sk_c)}\right)\ket{\Psi^{BCS}}=\left(v_ku_{-(k+sk_c)}-v_{-(k+sk_c)}u_k\right)c_{\downarrow,k+sk_c}^\dagger c_{\downarrow,-k}^\dagger\prod_{k'\neq -k,k+sk_c}\ket{\Psi^{BCS}_k}.
\end{align}
Therefore, we can evaluate the second order pertuabtive energy as 
\begin{align}
	\chi&=-\left(\dfrac{g\Omega}{4\delta}\right)^2\dfrac{1}{2N}\sum_{k,s}\dfrac{v_k^2u_{-(k+sk_c)}^2+u_k^2v_{-(k+sk_c)}^2-2u_kv_ku_{-(k+sk_c)}v_{-(k+sk_c)}}{\sqrt{\left(\dfrac{k^2}{2m}-\mu\right)^2+\Delta^2}+\sqrt{\left(\dfrac{(k+sk_c)^2}{2m}-\mu\right)^2+\Delta^2}}\\		
	E^{(2)}&=\chi(\alpha+\alpha^*)^2,
\end{align}
where $\chi$ is the susceptibility in Landau's paradigm. The Total Free energy of the system gives as 
\begin{align}
	F=\sum_{k}[(\dfrac{k^2}{2m}-\mu)-\mathcal{E}_k]-\dfrac{2N\Delta^2}{U}+\tilde{\omega}\left|\alpha\right|^2+\chi(\alpha+\alpha^*)^2+2\mu N+\mathcal{O}(\alpha^4),
\end{align}
where the higher order pertubative energy are contained in the term $\mathcal{O}(\alpha^4)$. The expression of the free energy refer to the category C, where the linear terms are absent in both sides. Therefore, all the four regions can occur in this case. But we can verify that $\dfrac{\partial^2F}{(\partial\Delta)^2}\Bigg|_{\Delta=0,\alpha=0}<0$ (actually for arbitrary $\alpha$) is always valid as $U<0$, so the region I is also absent. Thus we investigate the transition from region II into region IV, i.e., $(\Delta\neq0,\alpha=0)\rightarrow(\Delta\neq0,\alpha\neq0)$.

Consider the differentiation against the order parameter $\alpha^2$, the superradiant phase transition occurs at $\tilde{\omega}+4\chi=0$. Simultaneously, we also evaluate that $\dfrac{\partial^2F}{\partial\Delta\partial\left|\alpha\right|^2}\Bigg|_{\Delta=\Delta_0,\alpha=0}$ always give a positive value, indicating that the superradiance will suppress the pairing order.

Similar to the former example, we give linear fitting according to the second order differentiations, matching well with the numerial results shown in Fig. 2. 



\begin{figure}[tb]\label{1D_BCS_1st_PT}
	\centering
	\includegraphics[width=1.0\textwidth]{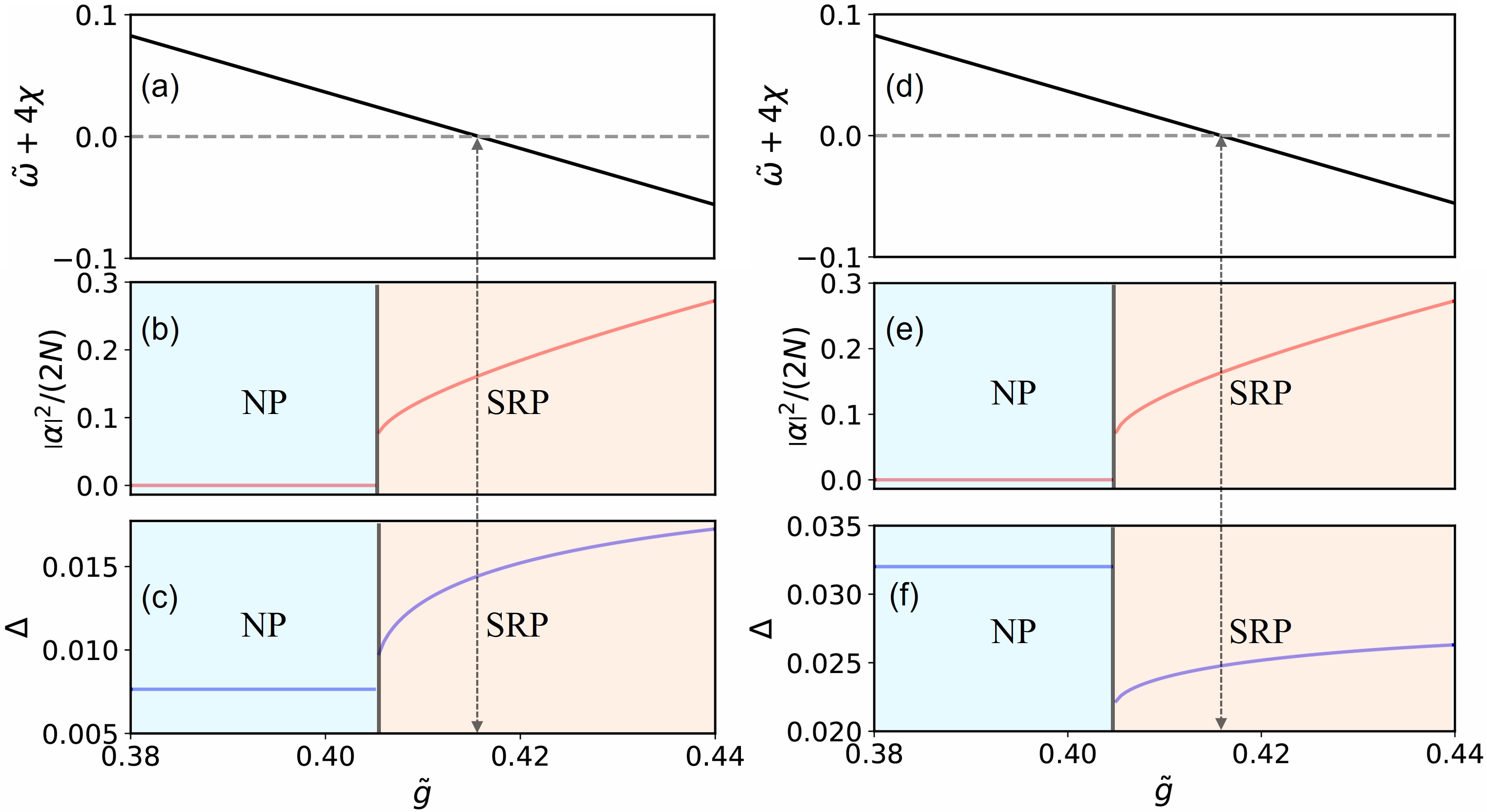}
	\caption{Here, we draw plots of the order parameters $\left|\alpha\right|^2$ and $\Delta$ change against the dimensionless couping strength $\tilde{g}$ with fixed filling factor $k_F=0.96$ and the attractive potential $U=-0.55$ in (a-c) and $U=-0.7$ in (d-f). This gives us the superfluid order and the critical coupling strength as $\Delta_0=0.0076,~\tilde{g}_c=0.405$, and $\Delta_0=0.032,~\tilde{g}_c=0.4045$, respectively. The black solid lines stand for the critcial boundaries seperate the region II ($\alpha=0, \Delta_0\neq0$) and region IV ($\alpha\neq0, \Delta_0\neq0$), which are colored as the same rule before. The black dashed arrow refers to the possible 2nd order critical point with $\tilde{\omega}+4\chi=0$.}
\end{figure}

Following the physical insights of previous researches, the 1D Fermionic Dicke model may give discontinuous phase transition in zero temperature limit~\cite{PhysRevLett.112.143003,vph8-186g} around $k_F\approx1$, thus we expect the simultaneous jumps of both the superfluid order $\Delta$ and the superradiant order $\alpha$. In Fig.~\ref{1D_BCS_1st_PT}, the filling factor is fixed as $k_F=0.96$ with different attractive potential $U=-0.55$ in (a-c) and $U=-0.7$ in (d-f) respectively. The subfigures (a) and (d) gives the $\tilde{\omega}+4\chi$ changing with the dimensionless coupling strength $\tilde{g}$, whose zero point leads to possible continuous phase transition. However, as is shown in (b)(c) and (d)(e), the phase transition occurs along with the sudden change of the order parameters, ahead of the disappearance of $\tilde{\omega}+4\chi$, which is marked by the dashed arrows. It turns out the 1st order (discontinuous) phase transition emerges. Meanwhile, the larger attractive potential $\left|U\right|$ will give larger pairing order $\Delta$ in normal phase ($\alpha=0$). But once entering the superradiance region, the pairing order will be more likely to be enhanced with smaller attractive potential.

Unfortunately, we can hardly give the 1st order critical coupling strength $\tilde{g}_c$, unless we can calculate all orders expressions of $\alpha$ and $\Delta$ (get the analytical expression of the free energy on $\alpha$ and $\Delta$). However, it provide us with the idea that the superfluid enhancing or suppression might be manipulated by the 1st order superradiant phase transition.


	\end{document}